\documentclass[a4paper,fleqn,usenatbib,useAMS]{mnras}

\usepackage{graphicx}	
\usepackage{amsmath}	
\usepackage{amssymb}	
\usepackage{multicol}        
\usepackage{bm}		
\usepackage{pdflscape}	
\usepackage{journals}

\usepackage[T1]{fontenc}
\usepackage{ae,aecompl}

\newcommand{\sbunit}{erg~s$^{-1}$~kpc$^{-2}$}

\newcommand{\hasb}{\ensuremath{\Sigma_\mathrm{H\alpha}}}

\newcommand{\kms}{\ensuremath{\mathrm{km~s^{-1}}}}

\newcommand{\ha}{{\rm H\ensuremath{\alpha}}}
\newcommand{\hb}{{\rm H\ensuremath{\beta}}}

\newcommand{\rtt}{\ensuremath{R\mathrm{_{23}}}}

\newcommand{\hii}{\hbox{H\,{\sc ii}}}

\newcommand{\oiil}{\hbox{[O\,{\sc ii}] $\lambda$3727}}

\newcommand{\oiiil}{\hbox{[O\,{\sc iii}] $\lambda$4959}}
\newcommand{\oiiir}{\hbox{[O\,{\sc iii}] $\lambda$5007}}
\newcommand{\oil}{\hbox{{\rm [O\,{\sc i}]}$\lambda$6300}}
\newcommand{\niir}{\hbox{[N\,{\sc ii}] $\lambda$6583}}
\newcommand{\niil}{\hbox{[N\,{\sc ii}] $\lambda$6548}}
\newcommand{\siil}{\hbox{[S\,{\sc ii}] $\lambda \lambda$6717, 6731}}

\newcommand{\oiii}{\hbox{[O\,{\sc iii}]}}

\newcommand{\oii}{\hbox{[O\,{\sc ii}]}}

\newcommand{\oi}{\hbox{{\rm [O\,{\sc i}]}}}
\newcommand{\nii}{\hbox{[N\,{\sc ii}]}}
\newcommand{\sii}{\hbox{[S\,{\sc ii}]}}

\def\lax{{$\mathrel{\hbox{\rlap{\hbox{\lower4pt\hbox{$\sim$}}}\hbox{$<$}}}$}}
\def\gax{{$\mathrel{\hbox{\rlap{\hbox{\lower4pt\hbox{$\sim$}}}\hbox{$>$}}}$}}

\def\kms{\hbox{km$\,$s$^{-1}$}}

\title[SDSS IV-MaNGA: Impact of DIG] {SDSS-IV MaNGA: The Impact of Diffuse Ionized Gas on Emission-line Ratios, Interpretation of Diagnostic Diagrams, and Gas Metallicity Measurements}

\author[K. Zhang et al.]{Kai~Zhang,$^{1}$\thanks{Contact e-mail: \href{mailto:zkdtckk@gmail.com}{zkdtckk@gmail.com}},
Renbin~Yan$^{1}$, Kevin~Bundy$^{2}$, Matthew~Bershady$^{3}$, L. Matthew~Haffner$^{3}$, 
\newauthor
Ren{\'e}~Walterbos$^{4}$, Roberto Maiolino$^{5,6}$, Christy~Tremonti$^{3}$, Daniel~Thomas$^{7}$, Niv~Drory$^{8}$, 
\newauthor
Amy~Jones$^{9}$, Francesco~Belfiore$^{5,6}$,  Sebastian~F. S{\'a}nchez$^{10}$,  
\newauthor
Aleksandar~M. Diamond-Stanic$^{3}$, Dmitry~Bizyaev$^{11,12}$, Christian~Nitschelm$^{13}$,  
\newauthor
Brett~Andrews$^{14}$, Jon~Brinkmann$^{11}$, Joel R. Brownstein$^{15}$,  Edmond~Cheung$^{2}$,  
\newauthor
Cheng~Li$^{16,17}$, David~R. Law$^{18}$, Alexandre~Roman Lopes$^{19}$,  Daniel~Oravetz$^{11}$, 
\newauthor
Kaike~Pan$^{11}$, Thaisa~Storchi-Bergmann$^{20, 21}$, Audrey~Simmons$^{11}$ \\
(Affiliations can be found after the references)
}

\pubyear{2016}

\begin{document}
\label{firstpage}
\pagerange{\pageref{firstpage}--\pageref{lastpage}}
\maketitle

\begin{abstract}

Diffuse Ionized Gas (DIG) is prevalent in star-forming galaxies. Using
a sample of 365 nearly face-on star-forming galaxies observed by
MaNGA, we demonstrate how DIG in star-forming galaxies impacts the
measurements of emission line ratios, hence the interpretation of
diagnostic diagrams and gas-phase metallicity measurements.  At fixed
metallicity, DIG-dominated low \hasb\ regions display enhanced
\sii/\ha, \nii/\ha, \oii/\hb, and \oi/\ha.  The gradients in these
line ratios are determined by metallicity gradients and \hasb.  In line
ratio diagnostic diagrams, contamination by DIG moves \hii\ regions
towards composite or LI(N)ER-like regions. A harder ionizing
  spectrum is needed to explain DIG line ratios. Leaky \hii\ region
models can only shift line ratios slightly relative to
\hii\ region models, and thus fail to explain the composite/LI(N)ER line
ratios displayed by DIG.  Our result favors ionization by evolved
stars as a major ionization source for DIG with LI(N)ER-like emission.

DIG can significantly bias the measurement of gas metallicity and
metallicity gradients derived using strong-line methods. Metallicities
derived using N2O2 are optimal because they exhibit the smallest bias
and error. Using O3N2, \rtt, N2=\nii/\ha, and N2S2\ha\ (Dopita et
al. 2016) to derive metallicities introduces bias in the
derived metallicity gradients as large as the gradient itself.

The strong-line method of Blanc et al. (2015; IZI hereafter) cannot be
applied to DIG to get an accurate metallicity because it currently
contains only \hii\ region models which fail to describe the DIG.

\end{abstract}


\section{Introduction}
\label{intro.sec}

Diffuse ionized gas (DIG hereafter) is an important gas component in
star-forming galaxies. It was first identified in our Milky Way (MW) off the disk, and known as the Reynolds layer
(Reynolds 1984). DIG is a major part of ionized gas in MW. In terms
of mass, it is about 30\% of the MW neutral hydrogen (Reynolds
1990, 1991). DIG is also found in external galaxies both in
extra-plannar halos (e.g., Dettmar 1990; Rand et al. 1990; Rand 1996;
Hoopes et al. 1999; Rossa \& Dettmar 2003a,b) and in the disk (e.g.,
Monnet 1971; Zurita et al. 2000; Oey et al. 2007). The contribution
of DIG to the total emission line flux for face-on galaxies is
substantial (e.g., Walterbos \& Braun 1994; Ferguson et al. 1996;
Hoopes, Walterbos, \& Greenawalt 1996; Greenawalt et al. 1998). For 109
star-forming galaxies in the SINGG sample, the DIG fraction in
\ha\ flux is 0.59$\pm$0.19, and this ratio depends on the \ha\ surface
brightness of the whole galaxy (Oey et al. 2007). DIG is important in
understanding the ionized gas in star-forming galaxies.

The differences between DIG and \hii\ regions are not only in emission
intensity, but also in emission line ratios, indicating different
physical conditions.  The \siil/\ha, \niil/\ha, (\sii/\ha\ and
\nii/\ha\ hereafter, e.g. Reynolds 1985a; Hoopes \& Walterbos 2003;
Madsen et al. 2006), \oil/\ha\ and \oiil/\hb\ (\oi/\ha\ and
\oii/\hb\ hereafter, e.g. Voges \& Walterbos 2006; Haffner et
al. 1999) are found to be enhanced in DIG relative to
\hii\ regions. More quantitively, \sii/\ha, \nii/\ha, \oii/\ha,
\oi/\ha\ ratios correlate negatively with \ha\ flux in the MW and
other galaxies (Reynolds et al. 1998; Haffner et al. 1999;
T{\"u}llmann et al. 2000; Hausen et al. 2002; Voges \& Walterbos 2006;
Madsen et al. 2006; Blanc et al. 2009).  In high spatial resolution
observations (pc to tens of pc, such as those by The Wisconsin H-Alpha Mapper (WHAM)) the line ratio
vs \hasb\ relation is a gradual transition from \hii\ regions to
DIG. In low spatial resolution observations (hundreds of pc or kpc,
such as integral-field studies of nearby galaxies) this relation is
due to the mixing of \hii\ regions and DIG within resolution elements.
Different fractions of DIG and \hii\ region contribution, combinned
with the different line ratios for these two types of regions would
naturally produce the trend we see. Observed relations between line
ratios and \hasb\ provide an empirical method to separate DIG
dominated and \hii\ region dominated regions (Blanc et al. 2009;
Kreckel et al. 2016; Kaplan et al. 2016).


DIG line ratios cannot be explained by models of \hii\ regions. On the
classical BPT diagram (Baldwin et al. 1981; Veilleux \& Osterbrock
1987; Kewley et al. 2001, 2006; Kauffmann et al. 2003), the location
of an \hii\ region is mainly determined by its metallicity and
ionization parameter (e.g., Kewley et al. 2002; Dopita et
al. 2013). DIG shows a lower ionization parameter than \hii\ regions, 
which explains partly if not mostly the enhancement of \nii/\ha, \sii/\ha, and \oii/\ha\ in DIG. However, just
varying these two parameters cannot fully produce the line ratios seen
in DIG (e.g., Galarza et al. 1999; Hoopes \& Walterbos 2003; Kaplan et
al. 2016). A third variable is needed. Also, the line ratios of DIG
{\it within} a galaxy often vary much more than those of
\hii\ regions. For example, in our Milky Way, DIG \sii/\ha\ ratios
display a dispersion of 0.13 (in linear space), compared with a
dispersion of 0.03 for \hii\ regions (Madsen et al. 2006). Temperature
has been proposed as the strongest factor explaining the variations of
DIG line ratios because it can explain the coherent variation of
\oiil/\ha, \niil/\ha, and \sii/\ha(Haffner et al. 1999; Mierkiewicz et
al. 2006). The variation of temperature is a result of the balance
between heating and cooling, which also requires a physical
explanation. Besides photoionization, an additional source of heating
might be important especially when the density is low (e.g., Reynolds
\& Cox 1992). We will show in this paper that a harder ionizing
spectrum could easily explain the DIG line ratios we observe in
star-forming galaxies. The harder spectrum is capable of producing
partially-ionized regions that emit strong \sii, \nii, \oii\ and \oi,
and an increase in temperature.
The DIG can impact our interpretation of the line-ratio diagnostic
diagrams of galaxies for either integrated or spatially-resolved
spectroscopy. The BPT
diagrams, for example, are widely used to diagnose the physical
properties of ionized gas and separate different types of galaxies
(Baldwin et al. 1981; Veilleux \& Osterbrock 1987; Kewley et al. 2001,
2006; Kauffmann et al. 2003 ). When metallicity, ionization parameter,
density, or ionizing spectrum change, the location of ionized gas on
the BPT diagram changes (e.g., Dopita et al. 2000, 2013; Kewley et
al. 2002, 2013a,b). When the N/O ratio is high, it is possible for a star-forming galaxy to be classified as
composite galaxy (P{\'e}rez-Montero \& Contini 2009; P{\'e}rez-Montero et al. 2013, 2016).  The enhanced
\nii/\ha\ and \sii/\ha\ of DIG would move the position of a
star-forming galaxy toward the composite or LI(N)ER region (e.g.,
Sarzi et al 2006; Stasi{\'n}ska et al. 2008; Yan \& Blanton 2012;
Kehrig et al. 2012; Singh et al. 2013; Gomes et al. 2016; Belfiore et al. 2016a,b) on the BPT
diagrams. Consequently, understanding DIG helps us understand the
nature of galaxies classified as composites or LI(N)ERs by the BPT diagrams.

The study of DIG is also critical for gas-phase metallicity
measurements in star-forming galaxies. Metallicity calibrations are
generally based on \hii\ region models which have certain
assumptions. Some of these assumptions are not valid for DIG, and
hence lead to biased metallicity measurements when DIG is present. The
{\it metallicity} and {\it ionization parameter}
(q=$\frac{ionizing\hspace{0.1cm}photon\hspace{0.1cm}flux}{Ne}=U\times
c$) determine the line ratios of a \hii\ region, and these two
parameters are correlated (Dopita et al. 2006). Metallicity also
determines the shape of the ionizing spectrum. In DIG, however, the
correlation between q and metallicity no longer holds, since DIG has a 
much lower ionization parameter than \hii\ regions, and the ionizing
spectrum shape can change. Given the very different line ratios such as
\nii/\ha, \oiiir/\hb\ (\oiii/\hb\ hereafter) and \sii/\ha\ for DIG and
\hii\ regions, biases in metallicities derived from strong line
methods are inevitable. These biases potentially contribute to the
large dispersion in metallicity measurements found in the
literature. This will influence metallicity gradient measurements
(S{\'a}nchez et al. 2014; Ho et al. 2015), metallicities at the
outskirts of galaxies (Moran et al. 2012), the mass-metallicity
relation in the local universe (e.g. McClure \& van den Bergh, 1968;
Lequeux et al. 1979; Garnett 2002; Tremonti et al. 2004; Lee et
al. 2006) and at high redshift (Erb et al. 2006; Maiolino et al. 2008;
Mannucci et al. 2009), and the mass-metallicity-SFR fundamental plane
(Mannucci et al. 2010; Yates et al. 2012; Andrews \& Martini 2013).
For single-fiber surveys such as SDSS, the observed emission lines
come from a combination of DIG and \hii\ regions, and the bias
introduced by DIG is uncertain and hard to quantify. With the help of
integral field spectroscopy (IFS), we can study how the presence of
DIG impacts metallicity measurements in detail.

In this paper, we demonstrate the prevalence of DIG in star-forming
galaxies from the MaNGA survey. The large sample and full optical
wavelength coverage enable us to explore the optical line ratios for
DIG for an unprecedented number of star-forming galaxies. 
  Section~\ref{sample.sec} describes our sample and the emission line
  measurements; while Section~\ref{s2.sec} shows the \sii/\ha\ vs
  \hasb\ relation that illustrates the dominance of DIG in low
  \hasb\ regions. Section~\ref{variation.sec} demonstrates how DIG
  impact line ratios like \sii/\ha, \nii/\ha, \oii/\hb, \oi/\ha,
  \oiii/\oii, \oiii/\hb. Section~\ref{diagnostic.sec} tests the leaky
  \hii\ region model as well as one in which hot evolved stars serve
  as the ionization source of the DIG, using different line ratios and
  diagnostic diagrams. Section~\ref{bias.sec} presents our study of
  how the DIG impacts metallicity derived using strong line methods:
  N2, N2O2, \rtt, O3N2 and N2S2\ha, and
  IZI. Section~\ref{discussion.sec} contains a discussion of these
  results, summarized then in Section~\ref{conclusion.sec}. We use a
cosmology with $H_{\rm 0}$ = 70 km\,s$^{-1}$\,Mpc$^{-1}$, $\Omega_{\rm
  m}$ = 0.3, and $\Omega_{\rm \Lambda}$ = 0.7 throughout this paper.

\section{Sample and Measurements}
\label{sample.sec}
\subsection{MaNGA Overview}

MaNGA (Mapping Nearby Galaxies at APO) (Bundy et al. 2015) is one of
the three core programs in the Sloan Digital Sky Survey-IV
(SDSS-IV). It aims at obtaining Integrated Field Spectroscopy (IFS) of
10,000 nearby galaxies. The survey employs the BOSS spectrographs
(Smee et al. 2013) on the 2.5m Sloan Foundation Telescope (Gunn et
al. 2006). The spectrographs provide a spectral coverage from
3600\AA\ to 10300\AA\ at a resolution around R$\sim$2000. The fiber
feed system has been re-designed (Drory et al. 2015) to accommodate
1423 fibers that are bundled into five 127-fiber bundles, two 91-fiber
bundles, four 61-fiber bundles, four 37-fiber bundles, two 19-fiber bundles,
twelve 7-fiber mini-bundles, and 92 sky fibers. All bundles with 19 or
more fibers are used for galaxy targets, while the mini-bundles are
used to observe standard stars providing a flux calibration to an
accuracy better than 5\% (Yan et al. 2016a). Each individual fiber is
2" in diameter. The filling factor of fibers in the bundles is
56\%. The observations are done with a dithering scheme to achieve a
complete spatial coverage and near critical sampling of the PSF (Law
et al. 2015).
  
The MaNGA sample is designed to have a roughly flat i-band absolute
magnitude distribution (Wake et al. in prep). The reconstructed
datacubes have 2.5" PSF and 0.5" spaxel (Law et al. 2015, 2016; Yan et
al. 2016a,b). The sample has a redshift range of $0.01<z<0.15$,
meaning the targets cover a factor of 15 in distance, and consequently
the physical resolution also span a range of 15. We select more
luminous galaxies, which are larger, at higher redshift. The typical
physical resolution is 1-2~kpc. The physical resolution is highly
correlated with luminosity which means that we need to be aware that
any results we find which might depend on physical resolution will be
specific to a specific range of luminosity and vice versa. The primary
sample, which covers to 1.5 effective radius ($R_e$) in major axis,
comprise 2/3 of the sample while the secondary sample, which covers to
2.5 $R_e$ comprise the remaining 1/3. Massive galaxies, which are
bigger, are selected at higher redshift. All plates are exposed
  until we reach a $(S/N)^2$ of 20 per pixel per fiber in the g-band
  continuum for a galactic-extinction-corrected g-band fiber magnitude
  of 22, and a $(S/N)^2$ of 36 per pixel per fiber in the i-band
  continuum for a galactic-extinction-corrected i-band fiber magnitude
  of 21 (Yan et al. 2016b). MaNGA provides a benchmark of resolved
ionized gas properties, stellar population, and dynamical evolution in
the local universe (e.g., Belfiore et al. 2015; Li et al. 2015;
Wilkinson et al. 2015).

\subsection{Sample Selection}

In this paper, we use a sample of 81 regular survey plates observed
before summer of 2015. There are 1391 unique galaxies observed. This
corresponds to the sample that was released in DR13. The
photometry data is from NASA-Sloan Atlas
catalog\footnote{http://www.nsatlas.org} (NSA Catalog). By applying a
color cut of $M_u - M_r<$2.1 ($M_u$ and $M_r$ are rest-frame absolute
magnitudes without extinction correction) we select only blue
galaxies. This resulted in a sample of 592 galaxies. To remove edge-on
galaxies we apply the cut $b/a>0.5$, where b and a are the minor and
major axis of the Sersic model. Edge-on galaxies suffer from strong
projection effects and are also prone to severe extinction. These
criteria leave us with 365 galaxies. AGNs are not eliminated
  because we care about DIG that is far away from the center. From
this sample we choose three galaxies that have large internal
variations of \ha\ surface brightness and best spatial resolution (127
fiber IFUs) to demonstrate the impact of DIG on line ratios,
interpretation of diagnostic diagrams, and metallicity
measurements. We discuss in Section~\ref{rainbow_ratio.sec} and
Section~\ref{rainbow_z.sec} that the impact of DIG is prevalent in all
star-forming galaxies in our sample.

\subsection{Spectrum Fitting}
\label{fitting.sec}

\begin{figure*}
\includegraphics[scale=0.45]{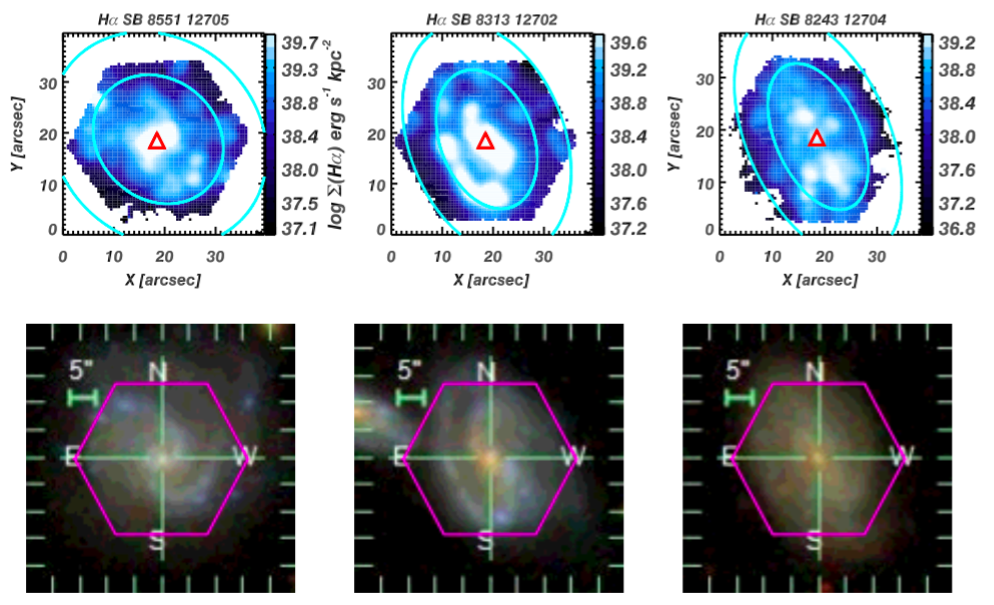} 
\caption{\ha\ surface brightness plot and SDSS optical image for each
  galaxy. The cyan ellipses in the upper represent are 1.5 and 2.5
  times the effective radius. The hexagons in the lower row
  demonstrate the location of the IFU bundles on the object. }
\label{hasb.fig}
\end{figure*}

We start with the datacubes produced by the MaNGA data reduction
pipeline (Law et al. 2016).  We first construrct Voronoi bins of the spectra by requiring
the S/N in r band in each bin to be greater than 30 per \AA. The
covariance between spaxels is not accounted for when binning. With
BC03\footnote{$http://www2.iap.fr/users/charlot/bc2003/$} (Bruzual \&
Charlot 2003), we produce 14 continuum templates for SSPs with
$Z_*$=0.02 and 0.008 and ages=13, 7, 2, 1, 0.5, 0.25, and 0.125Gyr.
For the stacked spectrum in each bin, we fit combinations of these
simple stellar populations (SSPs), fitting velocity and relative
amplitudes to spectra combined from spaxels in a Voronoi bin.  We then
derive the stellar velocity dispersion using the {\it vdispfit.pro} in
IDLSPEC2D\footnote{$http://spectro.princeton.edu/idlspec2d\_doc.html$}
package. 20\AA\ windows around \oiil, \hb, \oiiil, \oiiir, \oil,
  \niil, \ha, \niir, and \siil\ are masked during the continuum
  fitting. We smooth the observed spectrum with a 50 pixel window and
then subtract the smoothed spectrum from the original spectrum. This
smoothing basically removes remaining spectrophotometric calibration
mismatch between data and templates as well as effects of internal
extinction. The smoothing scale is $\sim$3000 \kms.  By doing this, we
are left with only small scale spectral variations such as absorption
lines and emission lines, which we refer to as the 'line feature
spectrum'. The same technique is applied to the SSP templates to get
the 'line feature templates' that only contains small scale
information. A linear regression is performed to fit the 'line feature
spectrum' with the sum of 'line feature templates' at a number of
velocity offsets relative to the systematic redshift of the
galaxy. The velocity grid ranges from -450 \kms\ to 450 \kms\ with an
interval of 30 \kms. For each Voronoi bin at each velocity, the
least-square fitting yields a $\chi^2$ for that fit. A quadratic curve
is fitted to the $\chi^2$ vs velocity curve to find the velocity
offset yielding the minimum $\chi^2$. For each spaxel in one Voronoi
bin, we use the fitted stellar continuum, the combination of 14 BC03
templates, as one template, and adjust the amplitude to fit the
continuum in each spaxel so that we can measure the emission-line
ratios on a spaxel by spaxel basis.  The residual emission-line only
spectrum is stored for refined emission line fitting.  We use single
gaussians to fit the \oi, \hb, \oiii, \niil, \ha, \niir, and
\sii\ respectively, and we use two gaussians to fit \oii. The line
ratio of \niir/\niil\ is fixed to 3. Velocity dispersions and
  central velocity shifts of all emission lines are tied to be the
  same. The \ha\ surface brightness maps for the 3 representative
galaxies are shown in Figure~\ref{hasb.fig} .
The errors of the line strength are obtained by the MPFIT package
which only includes the fitting errors (Markwardt 2009). A S/N
  cut of 5 is applied to emission lines used. 

\section{Variation of Line Ratios with \ha\ Surface Brightness}
\label{variation.sec}

\subsection{Separation of regions dominated by DIG and \hii\ regions }
\label{s2.sec}
\begin{figure*}
\includegraphics[scale=1.0]{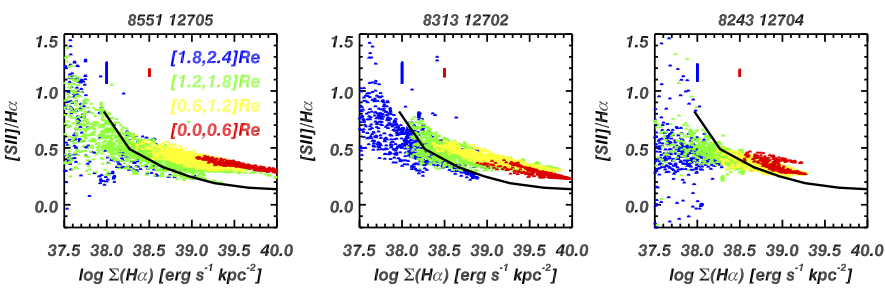}
\caption{\sii/\ha\ vs \ha\ surface brightness plot for each
  galaxy. Different color represent different annuli. The solid line
  is the \sii/\ha\ vs \ha\ surface brightness relation for Milky Way
  (Madsen et al. 2006). We see that high and low \ha\ surface
  brightness regions have different \sii/\ha\. The red and blue bars
  show the typical error at log \hasb=38.5 \sbunit\ and log \hasb=38
  \sbunit\ for individual spaxels. The length of the error bars is
  2$\sigma$. }
\label{s2.fig}
\end{figure*}

The first step to study DIG is separating DIG from \hii\ regions. The
best way to isolate DIG is to identify individual \hii\ regions and
subtract them out (e.g., Walterbos \& Braun 1994; Zurita et al. 2000,
2002; Thilker, Walterbos, \& Braun 2002; S{\'a}nchez et al. 2012). Due
to the limited spatial resolution of MaNGA, we can not resolve
individual \hii\ regions in one galaxy. Our reconstructed PSF is about
2.5" FWHM (2" covers 1~kpc at z=0.025) while the size of a typical
\hii\ region is a few to hundreds of $pc$ (Kennicutt 1984; Garay \&
Lizano 1999; Kim \& Koo 2001; Hunt \& Hirashita 2009). So the light in
one spaxel is always a mixture of \hii\ region emission and the
surrounding DIG. Instead we separate DIG dominated regions and
\hii\ regions dominated regions. In Figure \ref{s2.fig}, we see that
the low surface brightness regions have \sii/\ha$\sim0.5-1.0$ while
the high surface brightness regions show \sii/\ha$\sim0.2-0.4$,
depending on the metallicity. Since \sii/\ha\ is sensitive to
metallicity, and there are metallicity gradients in galaxies, we
separate all the spaxels in each galaxy into different radial bins:
[0,0.6],[0.6,1.2], [1.2,1.8], and [1.8,2.4]$R_e$.  The \sii/\ha\ vs
\hasb\ relations are similar in all radial bins, meaning the variation
of \sii/\ha\ is not caused by metallicity variation, but reflects
the transition from DIG dominated low \hasb\ regions to \hii\ region
dominated high \hasb\ regions. {This figure illustrates that \hasb\ can be
  used to separate the two different kinds of regions: low \hasb\ DIG
  dominated regions and high \hasb\ \hii\ region dominated regions. }

\subsection{\sii/\ha, \nii/\ha, \oii/\hb, and \oi/\ha }
\label{PIR.sec}

Galaxies show metallicity variations, in particular in the form of
radial gradients (e.g. S{\'a}nchez et al. 2014; Ho et al. 2015). It is
crucial to control metallicity before exploring how the line ratios
vary with other parameters like \ha\ surface brightness because
metallicity is a major source of variation in line ratios.  We plot
line ratios as a function of radius, and color code the dots with
\ha\ surface brightness. We assume that the metallicity is constant
within annuli for each galaxy but may be changing with radius.

\begin{figure*}
\includegraphics[scale=0.55]{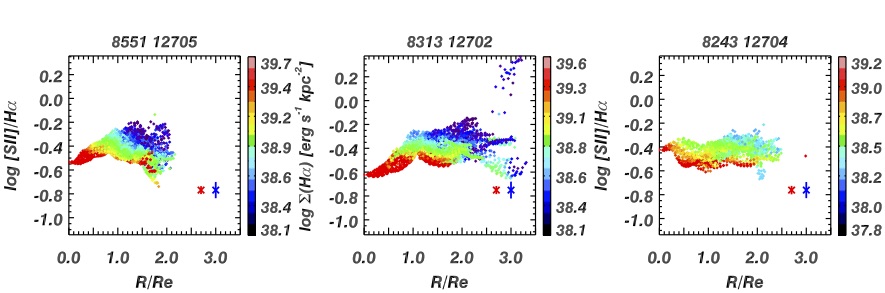} 
\caption{ \sii/\ha\ as a function of radius. The dispersion at fixed
  radius is about 0.2~dex. But when the dots are color-coded by \ha
  surface brightness as shown in the colorbar, we see a beautiful
  rainbow pattern. The dispersion is significantly reduced at fixed
  \hasb. This is because DIG that dominates the low \hasb\ region has
  higher \sii/\ha\ . The red and blue bars show the typical line ratio
  error at log\hasb=39 \sbunit\ and log\hasb=38.5 \sbunit\ for
  individual spaxels. The length of the error bars is 2$\sigma$. We
  show in Figure~\ref{rainbow_ratio.fig} that the impact of DIG is
  prevalent in all star-forming galaxies in our sample, not only
  in the three galaxies we show.  }
\label{s2_gradient.fig}
\end{figure*}

We explore the impact of DIG on \sii/\ha\ first. We color-code
  the dots by \ha\ surface brightness and show how \sii/\ha\ changes
  with radius in Figure \ref{s2_gradient.fig}. At a fixed
  radius, we see a rainbow pattern such that low \hasb\ regions have
  higher \sii/\ha. This is a direct demonstration that \sii/\ha\ is
  enhanced in DIG after controlling for metallicity. 

\begin{figure*}
\includegraphics[scale=0.55]{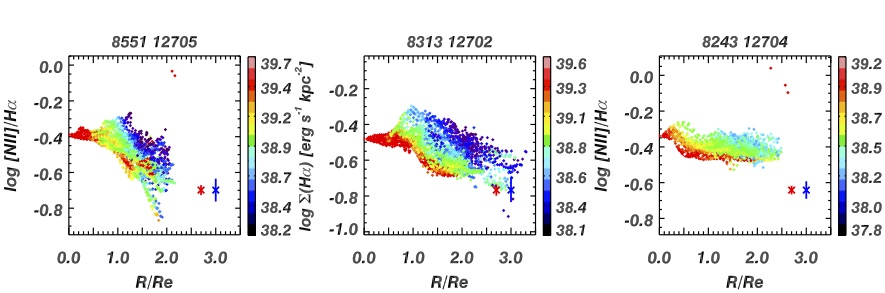} 
\caption{ \nii/\ha\ as a function of radius. We see
  \nii/\ha\ generally drops toward large radius due to a metallicity
  gradient. The dispersion at fixed radius is about 0.2~dex. But when
  the dots are color-coded by \ha\ surface brightness as shown in the
  colorbar, we see a beautiful rainbow pattern. This shows
  \nii/\ha\ is enhanced in DIG. The red and blue bars show the typical
  line ratio error at log\hasb=39 \sbunit\ and log\hasb=38.5
  \sbunit\ for individual spaxels. The length of the error bars is
  2$\sigma$. }
\label{n2.fig}
\end{figure*}

We then explore how DIG impacts \nii/\ha. \nii/\ha\ is used as a
metallicity indicator since Nitrogen is a secondary element and
proportional to $Z^2$ while \ha\ is not sensitive to metallicity
(e.g. Storchi-Bergmann et al., 1994, van Zee et al., 1998,
Denicol{\'o} et al., 2002).  We color-code the dots by \ha\ surface
brightness and show how \nii/\ha\ changes with radius in Figure
\ref{n2.fig}.  If we control for \ha\ surface brightness by looking at the
dots with the same color, \nii/\ha\ decreases towards large radius,
reflecting a metallicity gradient. Due to the presence of an
\ha\ surface brightness gradient, the different surface brightness
bins usually trace different parts of the galaxy. However, at the
radius where they overlap, we see a rainbow pattern. At a fixed
radius, DIG dominated low surface brightness region show a higher
\nii/\ha. If we use N2 to derive the metallicity, the enhancement
means that the metallicity would be overestimated in those spaxels
with a high DIG fraction and vice-versa.  The impact of DIG on the
metallicity derived using \nii/\ha\ is given in Section~\ref{z_n2.sec}.

\begin{figure*}
\includegraphics[scale=0.55]{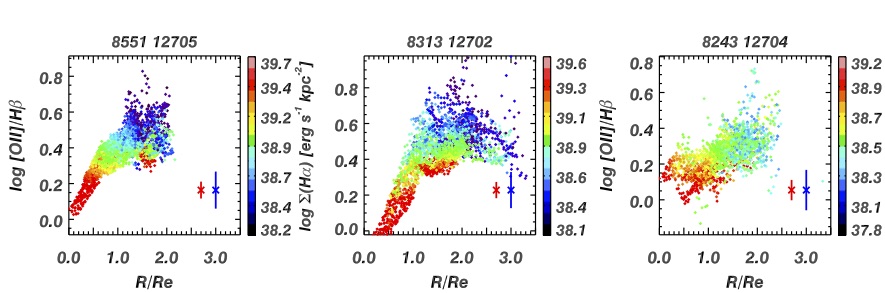} 
\caption{ \oii/H$\beta$ as a function of radius. The dots are
  color-coded by \hasb. We see rainbow patterns as in
  Figure~\ref{n2.fig}, indicating \oii/H$\beta$ is enhanced in
  DIG. The red and blue bars show the typical line ratio error at
  log\hasb=39 \sbunit\ and log\hasb=38.5 \sbunit\ for individual
  spaxels. The length of the error bars is 2$\sigma$.  }
\label{o2.fig}
\end{figure*}
\begin{figure*}
\includegraphics[scale=0.55]{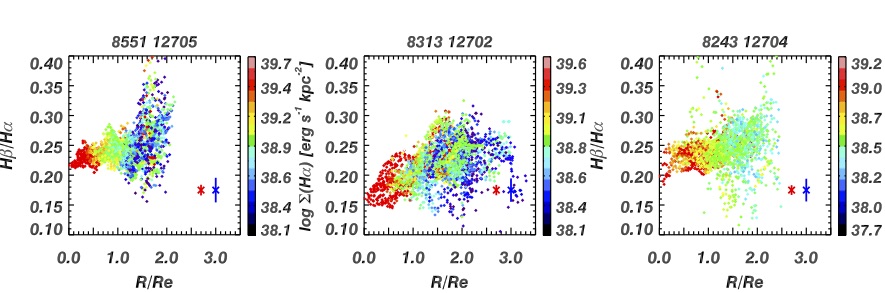} 
\caption{H$\beta$/\ha\ as a function of radius. The dots are
  color-coded by \hasb. We don't see any dependence of
  H$\beta$/\ha\ on \hasb\ at a given radius, implying the extinction
  correction is not the reason for the different line ratios of DIG
  and \hii\ regions. The red and blue bars show the typical line ratio
  error at log\hasb=39 \sbunit\ and log\hasb=38.5 \sbunit\ for
  individual spaxel. The length of the error bars is 2$\sigma$.  }
\label{hba.fig}
\end{figure*}

We show how \oii/\hb\ changes with radius and \ha\ surface brightness
in Figure \ref{o2.fig}. The extinction correction is not applied
because it is not reliable when the emission line, especially \hb, is
weak. Besides, a foreground dust screen may not be the appropriate
model for DIG.  Finally, We show in Figure~\ref{hba.fig} that
\hb/\ha\ does not depend on \ha\ surface brightness, meaning
extinction will not produce any line ratios change between DIG and
\hii\ regions.

[O\,{\footnotesize II}]/\hb\ generally increases with radius at fixed
\ha\ surface brightness due to a metallicity gradient. At fixed
radius, the low \ha\ surface brightness regions have higher
\oii/\hb. It is interesting that \oii/\hb\ and \nii/\ha\ both increase
with decreasing surface brightness at fixed radius while they change
reversely with radius at fixed \ha\ surface brightness. The opposite
variation trends of \oii/\hb\ and \nii/\ha\ with radius at fixed
surface brightness are consistent with a metallicity variation.
 With [O/H] above solar value, \nii/\ha\ increases with metallicity
due to the addition of secondary Nitrogen while \oii/\hb\ drops with
increasing metallicity because of decreasing temperature. The positive
correlation between \oii/\hb\ and \nii/\ha\ with surface brightness at
fixed radius is seemingly consistent with temperature variation.
Mierkiewicz et al. (2006) show that \oii/\ha\ and \nii/\ha\ correlate
positively with temperature variation and they concluded that the
variation of line ratios is driven by temperature variations (Haffner
et al. 1999; Haffner et al. 2009). However, the variation of
temperature is a result of the balance between heating and cooling
which, itself, needs a physical explanation. Besides, some signficant
assumptions need to be made to use \oii/\hb\ and \nii/\ha\ as
temperature tracers. The biggest assumption is that \oii\ and \hb\ are
co-spatial. However, we only resolve galaxies to kpc scale for MaNGA
survey, thus we see the integrated emission from all layers of ISM
around an ionizing source. The co-spatial assumption may not hold for
DIG. \nii/\ha\ line ratios are often used as a metallicity indicator for
extra-galactic studies, while it may be used as a temperature indicator
for galactic uses. Both metallicity and temperature are relevant to
\nii/\ha\ emission. Finally, a decrease in ionization parameter
  and/or a harder ionizing spectrum towards lower \hasb\ will also
  translate into enhancement of \oii/\hb\ and \nii/\ha. This is
  further explored in Section~\ref{diagnostic.sec}. 

\begin{figure*}
\includegraphics[scale=0.55]{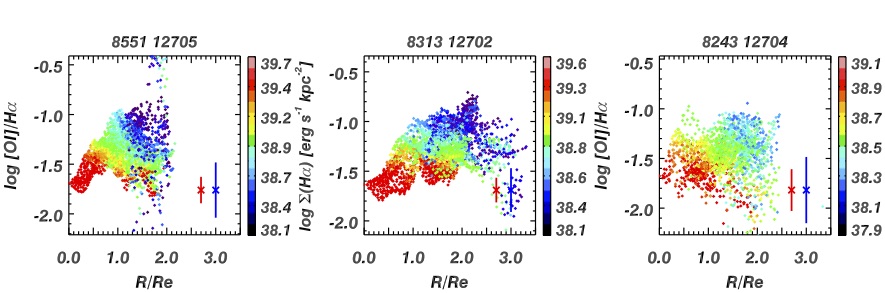} 
\caption{ \oi/\ha\ as a function of radius. The dots are color-coded
  by \hasb. We see rainbow patterns as in Figure~\ref{n2.fig},
  indicating \oi/\ha\ is enhanced in DIG. The red and blue bars show
  the typical line ratio error at log\hasb=39 \sbunit\ and
  log\hasb=38.5 \sbunit\ for individual spaxels. The length of the
  error bars is 2$\sigma$. }
\label{o1.fig}
\end{figure*}

[O\,{\footnotesize I}] is an important line to diagnose the physical
properties of ionized gas (Dopita et al. 2000; Kewley et
al. 2006). \oi\ is detected in DIG of M33 and \oi/\ha\ is found to
correlate negatively with emission measure (Voges \& Walterbos 2006).
\oi/\ha\ is a strong function of temperature and the hardness of the
ionizing spectrum. \oi\ has a high critical density
($n_{[OI]}=10^{6.3} cm^{-3}$), thus it is emitted in high density
neutral and partially ionized regions. \oi/\ha\ is also an
  excellent tracer of shocks (Dopita et al. 2000; Kewley et al. 2002;
  Allen et al. 2008). In Figure \ref{o1.fig}, we show how the
\oi/\ha\ changes with \ha\ surface brightness and radius for our three
galaxies. For all galaxies, \oi/\ha\ increases with radius. At a fixed
radius, \oi/\ha\ is enhanced for low surface brightness regions,
indicating \oi/\ha\ is higher in DIG. We note the contrast between DIG
and \hii\ region is large, \oi/\ha\ drops $\sim1~dex$ with an increase
of 1~dex in \hasb.

\subsection{\oiii/\oii\ and \oiii/\hb }
\label{o3.sec}
\begin{figure*}
\includegraphics[scale=0.55]{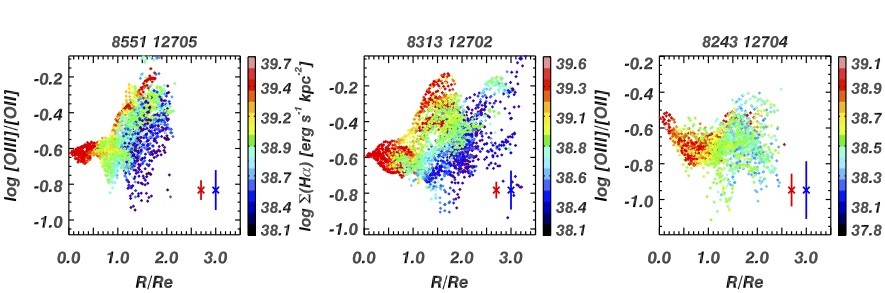} 
\caption{ \oiii/\oii\ as a function of radius. The dots are color-coded
  by \hasb. At a given radius, DIG shows lower \oiii/\oii, meaning
  lower ionization parameter. The red and blue bars show the typical
  line ratio error at log\hasb=39 \sbunit\ and log\hasb=38.5
  \sbunit\ for individual spaxels. The length of the error bars is
  2$\sigma$.  }
\label{o3o2.fig}
\end{figure*}

[O\,{\footnotesize III}]/\oii\ is a good ionization parameter proxy
when metallicity is controlled (Kewley et al. 2002, see also
Figure~\ref{grids.fig}).  We show how the \oiii/\oii\ values of low surface
brightness spaxels differ from the high surface brightness spaxels for
our sample galaxies in Figure~\ref{o3o2.fig}. This relation has large
dispersion but there is a clear trend. At a fixed \ha\ surface
brightness, \oiii/\oii\ increases as we go to larger radii due to a
metallicity gradient (see also Figure~\ref{o3o2n2o2.fig} ). At a fixed
radius, \oiii/\oii\ mostly decreases with decreasing surface
brightness. The result suggests that DIG has a lower ionization
parameter than \hii\ regions. 

[O\,{\footnotesize III}]/\hb\ is another frequently used ionization
parameter proxy. Its dependence on metallicity and ionizing spectrum
harndess is much stronger than \oiii/\oii.  How \oiii/\hb\ of DIG
differs from that of \hii\ regions depends on the specific physical
properties of the ISM. DIG can show either higher (Wang, Heckman, \&
Lehnert 1997; Rand 1998, 2000; Collins \& Rand 2001; Otte 2001, 2002;
Otte, Gallagher, \& Reynolds 2002) or lower (MW: Reynolds
1985b; M31: Greenawalt et al. 1997; Galarza, Walterbos, \& Braun 1999)
\oiii/\hb\ than \hii\ regions.
The study of our whole sample of 395 galaxies in
Section~\ref{rainbow_ratio.sec} confirms the diverse behavior of
\oiii/\hb\ in DIG. In the previous paragraph we have shown that DIG
has lower ionization parameter than \hii\ regions. At a fixed
  metallicity, a decrease in ionization parameter will result in a
  decease of \oiii/\hb\ for q<7.5 if no other parameter in the
  photoionization model is changed (see
  Figure~\ref{o3hbn2ha.fig}). For q>7.5, \oiii/\hb\ is roughly
  constant and independent of ionization parameter. This diverse
behavior of \oiii/\hb\ means a third parameter other than metallicity
and ionization parameter is needed, as will be discussed in
Section~\ref{diagnostic.sec}.
 The 3rd galaxy: 8243-12704 shows a
  negative \oiii/\oii\ and \oiii/\hb\ gradient. This is not consistent
  with an inverse metallicity gradient at the center because the
  \nii/\oii\ gradient is negative in this range. This could be due to
  AGN activity, but the strength of the AGN is not strong enough to
  produce AGN-like line ratios on the BPT diagram. These kind of
  sources will be explored in future work.

\begin{figure*}
\includegraphics[scale=0.55]{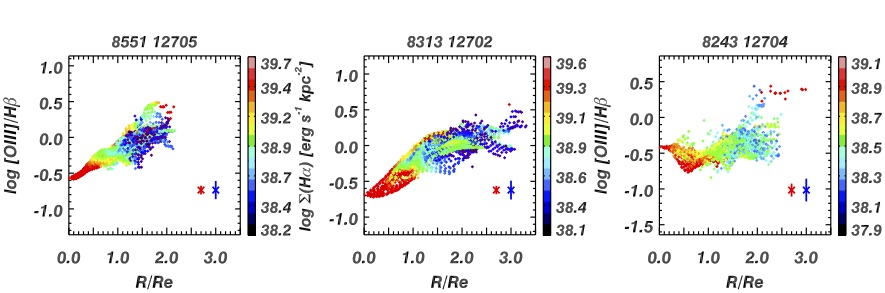} 
\caption{\oiii/H$\beta$ as a function of radius. The dots are
  color-coded by \hasb. At a given radius, DIG might show higher or
  lower \oiii/H$\beta$ than \hii\ regions depending on the specific
  situation of a galaxy. The red and blue bars show the typical line
  ratio error at log\hasb=39 \sbunit\ and log\hasb=38.5 \sbunit\ for
  individual spaxels. The length of the error bars is 2$\sigma$.  }
\label{o3.fig}
\end{figure*}


\subsection{Line ratios vs \hasb\ relation}
\label{rainbow_ratio.sec}
\subsubsection{Line ratios vs \hasb\ relation: The Whole Sample}
\begin{figure*}
\includegraphics[scale=0.55]{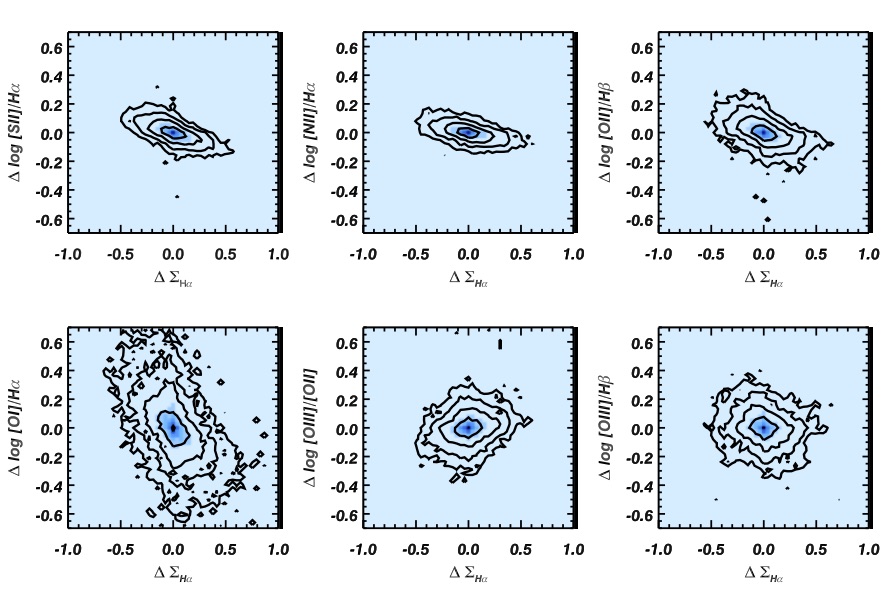} 
\caption{ $\Delta$ log (line ratios) vs $\Delta$ log \hasb\ relation
  at [0.4$R_e$, 0.6$R_e$] for all galaxies. We normalize the log line
  ratios vs log \hasb\ relation by subtracting the median log line
  ratio and median log \hasb. All the spaxels in a galaxy are weighted
  by $\frac{1}{N}$, where N is the number of valid spaxels in this
  galaxy. Contour levels are 5, 40, 70, 85, and 92
  percentile. \sii/\ha, \nii/\ha, \oii/\hb, \oi/\ha\ are higher in DIG
  dominated low \hasb\ regions. \oiii/\oii\ is slightly lower in DIG
  while \oiii/\hb\ is mildly higher in DIG.  }
\label{rainbow_ratio_all.fig}
\end{figure*}

 To quantify the variation of the line ratios as a function of
  \ha\ surface brightness for the whole sample, we select all spaxels
  that have [0.4$R_e$, 0.6$R_e$] in each galaxy. We then normalize the
  log line ratios vs log \hasb\ relation of each galaxy by subtracting
  from all the spaxels the median log line ratio and the median log
  \hasb\ of the set of spaxels. All the spaxels in a galaxy are
  weighted by $\frac{1}{N}$, where N is the number of valid spaxels in
  this galaxy. Figure~\ref{rainbow_ratio_all.fig} shows $\Delta$ log
  line ratios vs $\Delta$ log \hasb\ relation for the whole sample by
  combining all galaxies. \sii/\ha, \nii/\ha, \oii/\hb, \oi/\ha\ are
  higher in DIG dominated low \hasb\ regions. \oiii/\oii\ is slightly
  lower in DIG while \oiii/\hb\ is mildly higher in DIG.

\begin{figure*}
\includegraphics[scale=0.55]{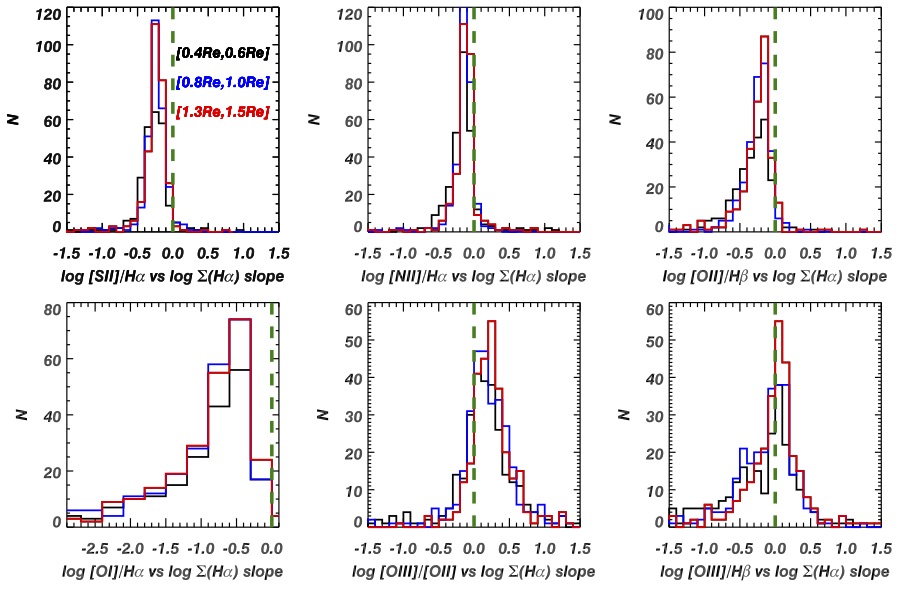} 
\caption{The distribution of $\Delta$ log line ratios vs $\Delta$ log
  \hasb\ slope at three radii. The black line is for the radius at
               [0.4$R_e$, 0.6$R_e$], the blue line is for the radius
               at [0.8$R_e$, 1.0$R_e$], and the red line is for the
               radius at [1.3$R_e$, 1.5$R_e$]. The green dashed line
               is the slope=0 line for reference. For almost all
               star-forming galaxies in our sample, DIG shows higher
               \sii/\ha\, \nii/\ha\, \oii/H$\beta$, \oi/\ha\, and
               lower \oiii/\oii\ than \hii\ regions.  \oiii/H$\beta$
               in DIG can be higher or lower than in \hii\ regions. }
\label{rainbow_ratio.fig}
\end{figure*}

Furthermore, we perform linear regression to the line ratio vs
\ha\ surface brightness relation in narrow annuli for individual
galaxies. For each galaxy, we get the slope of the linear regression at
three radial bins: [0.4$R_e$, 0.6$R_e$], [0.8$R_e$, 1.0$R_e$], and
[1.3$R_e$, 1.5$R_e$]. The distribution of slopes for different line
ratios at three radii are shown in black, blue, and red in
Figure~\ref{rainbow_ratio.fig}. A negative slope means the line ratio
is enhanced in DIG while a positive slope means the line ratio is
lower in DIG. We see that the slope peaks around $\sim$-0.3 for log
\sii/\ha, log \nii/\ha\, and log \oii/\hb. The slopes for these three
ratios are rarely positive. The slope peaks at -1.0 for log \oi/\ha,
which is the most significant among the line ratios explored here. The
slopes for \oiii/\oii\ is mostly positive, and its dispersion is
larger than the dispersion for \sii/\ha, \nii/\ha, and \oii/\hb. The
larger dispersion tells us that the ionization parameter varies
significantly.
The slope distribution for $\Delta$ log \oiii/\hb\ vs $\Delta$ log
\hasb\ peaks at $\sim$0, but the distribution is skewed to
negative values. There are more galaxies with higher \oiii/\hb\ in DIG than
galaxies with lower \oiii/\hb\ in DIG. The dispersion is even larger
than that for \oiii/\oii. This illustrates that \oiii/\hb\ is very
diverse in DIG. As shown in Section~\ref{o3.sec}, DIG shows higher
\oiii/\hb\ than \hii\ regions in some galaxies, while it shows lower
or similar \oiii/\hb\ for other
galaxies. Figure~\ref{rainbow_ratio.fig} demonstrates that DIG with
higher \sii/\ha, \nii/\ha, \oii/\hb, \oi/\ha\ and lower
\oiii/\oii\ than \hii\ regions is prevalent among all galaxies in our
sample. \oiii/\hb\ could be significantly higher or lower than
\hii\ regions. The contamination of DIG influences every star-forming
galaxy.

\subsubsection{Line ratios vs \hasb\ relation: Split by Stellar Mass}
\begin{figure*}
\includegraphics[scale=0.55]{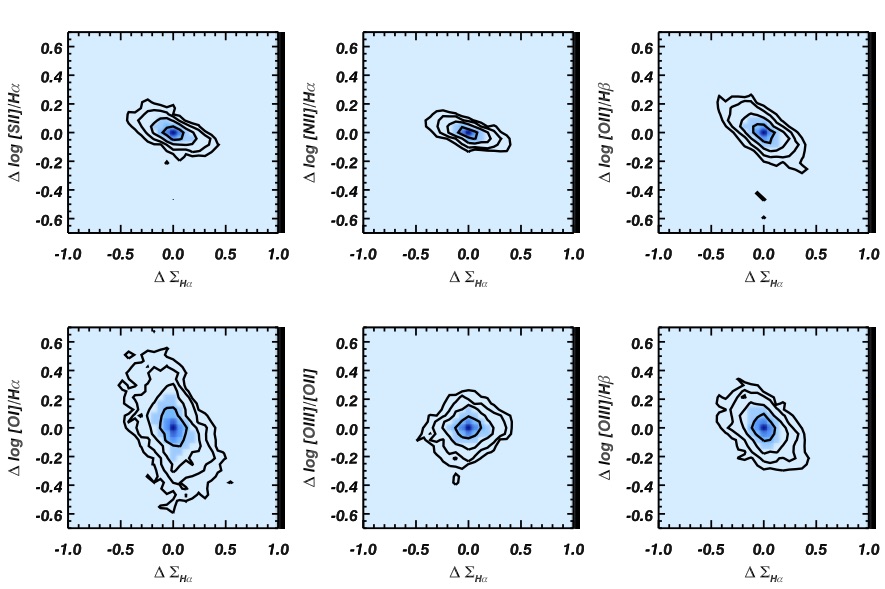} 
\caption{The $\Delta$ log (line ratios) vs $\Delta$ log \hasb\ relation
  at [0.4$R_e$, 0.6$R_e$] for the one third most massive galaxies. We
  normalize the log line ratios vs log \hasb\ relation by subtracting
  the median log line ratio and median log \hasb. All the spaxels in a
  galaxy are weighted by $\frac{1}{N}$, where N is the number of valid
  spaxels in the galaxy. \sii/\ha, \nii/\ha, \oii/\hb, \oi/\ha\ are
  higher in DIG dominated low \hasb\ regions. \oiii/\oii\ is slightly
  lower in DIG while \oiii/\hb\ is significantly higher in
  DIG. \oiii/\hb\ in DIG of massive galaxies is different from that in
  less massive galaxies.  }
\label{rainbow_ratio_Mstarhigh.fig}
\end{figure*}

\begin{figure*}
\includegraphics[scale=0.55]{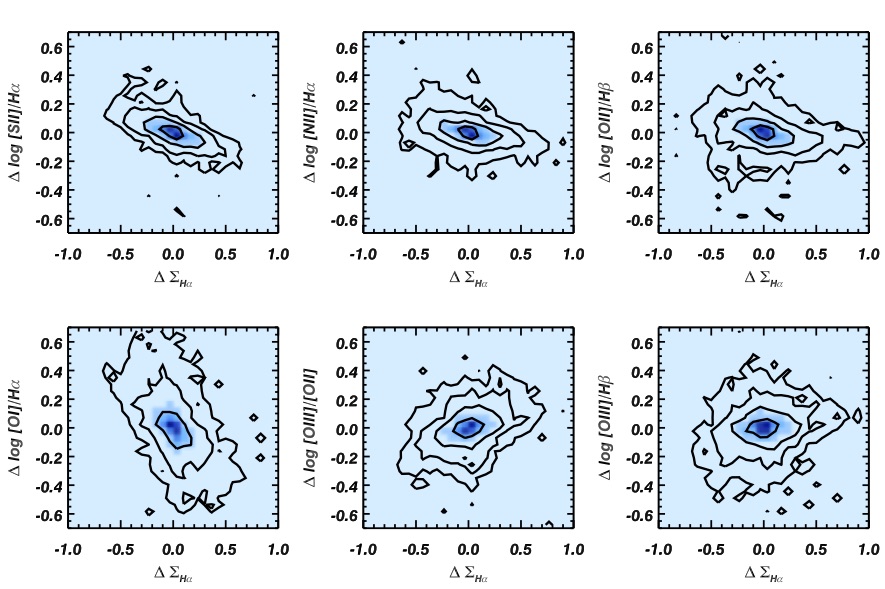} 
\caption{The $\Delta$ log (line ratios) vs $\Delta$ log \hasb\ relation
  at [0.4$R_e$, 0.6$R_e$] for the one third least massive galaxies. We
  normalize the log line ratios vs log \hasb\ relation by subtracting
  the median log line ratio and median log \hasb. All the spaxels in a
  galaxy are weighted by $\frac{1}{N}$, where N is the number of valid
  spaxels in the galaxy. \sii/\ha, \nii/\ha, \oii/\hb, \oi/\ha\ are
  higher in DIG dominated low \hasb\ regions. \oiii/\oii\ is slightly
  lower in DIG while \oiii/\hb\ is slightly lower in DIG.  }
\label{rainbow_ratio_Mstarlow.fig}
\end{figure*}

 In Figure~\ref{rainbow_ratio_Mstarhigh.fig} and
  Figure~\ref{rainbow_ratio_Mstarlow.fig}, we show the $\Delta$ log
  line ratio vs $\Delta$ log \hasb\ relations for galaxies with
  stellar mass less than $10^{9.43}$ and higher than $10^{10.08}$
  respectively. These are the one third least massive and the one
  third most massive galaxies in our sample. The slopes of $\Delta$
  log \sii/\ha, \nii/\ha, \oii/\hb, \oi/\ha, and \oiii/\oii\ vs
  $\Delta$ log \hasb\ relations do not depend much on stellar
  mass. However, massive galaxies show a significantly negative
  $\Delta$ log \oiii/\hb\ vs $\Delta$ log \hasb\ relation while less
  massive galaxies show a positive $\Delta$ log \oiii/\hb\ vs $\Delta$
  log \hasb\ relation. In other words, \oiii/\hb\ of DIG is always
  enhanced relative to \hii\ regions in the most massive galaxies. A
  leaky \hii\ region model can not produce high \oiii/\hb\ relative to
  \hii\ regions (Section~\ref{grids.sec}). The dependence of the $\Delta$
  log \oiii/\hb\ vs $\Delta$ log \hasb\ relation on stellar mass may
  indicate different ionization mechanisms in galaxies of different
  masses. We leave the study of the physical reason behind the
  dependence of $\Delta$ log line ratios vs $\Delta$ log
  \hasb\ relation on stellar mass for future studies. 

\section{Diagnostic Diagrams}
\label{diagnostic.sec}
\subsection{Comparison of line ratios with \hii\ region grids}
\begin{figure*}
\includegraphics[scale=0.55]{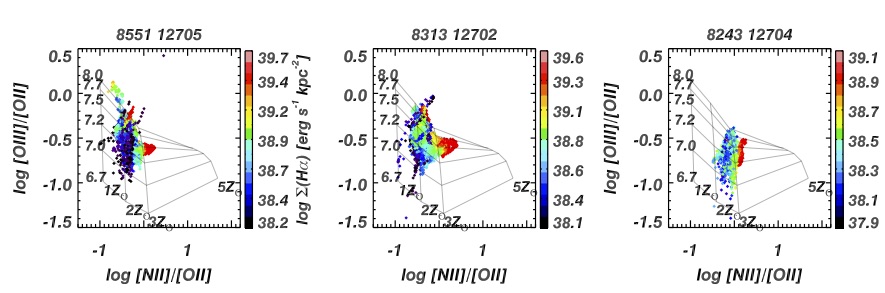} 
\caption{\oiii/\oii vs \nii/\oii\ diagram (Dopita et al. 2000,
  2013). The grids are from Dopita et al. (2013) with
  $\kappa=Inf$. The nearly-horizontal lines are for constant
  ionization parameters, and the nearly vertical lines are for
  constant metallicities. The labels denote the metallicities and
  ionization parameters (log q) for the grid.  }
\label{o3o2n2o2.fig}
\end{figure*}

\begin{figure*}
\includegraphics[scale=0.55]{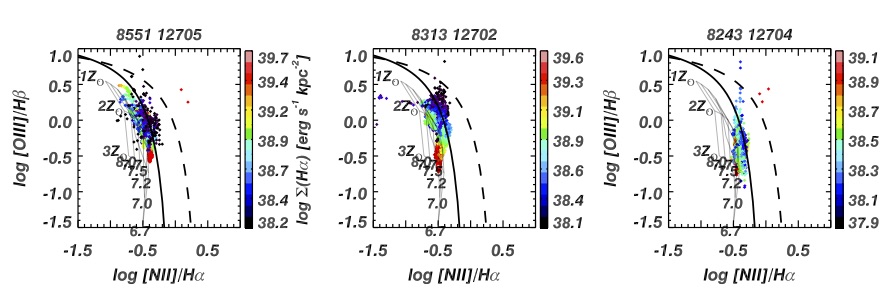} 
\caption{\oiii/H$\beta$ vs \nii/\ha\ diagram for each
  galaxy. Different colors represent different $\Sigma_{\ha}$. The
  solid and dashed lines are the demarcation line from Kauffmann et
  al. (2003) and Kewley et al. (2001). The grids are from Dopita et
  al. (2013) with $\kappa=Inf$. DIG can not be covered by the grids,
  suggesting HII region models with variations of metallicity and
  ionization parameter can not produce the DIG line ratios.  }
\label{o3hbn2ha.fig}
\end{figure*}

\begin{figure*}
\includegraphics[scale=0.55]{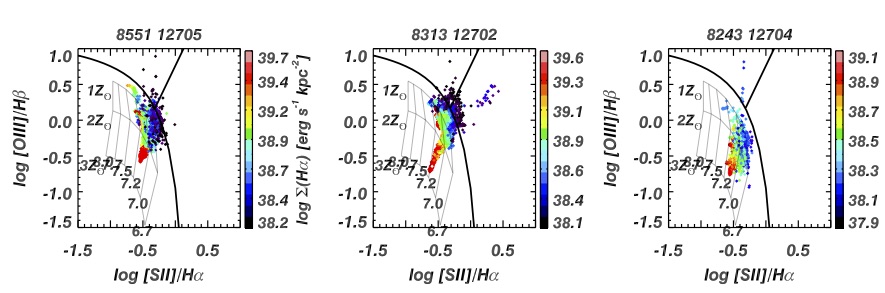} 
\caption{\oiii/H$\beta$ vs \sii/\ha\ diagram for each
  galaxy. Different colors represent different $\Sigma_{\ha}$. The
  solid lines are the demarcation lines to separate Seyfert galaxies
  from LINER in Kewley et al. (2006). The grids are from Dopita et
  al. (2013) with $\kappa=Inf$. The grids do not cover DIG, suggesting
  HII region models with variations of metallicity and ionization
  parameter can not produce the DIG line ratios.}
\label{o3hbs2ha.fig}
\end{figure*}

In Figure \ref{o3o2n2o2.fig}, we plot the line ratios for each spaxel 
on the \oiii/\oii\ vs \nii/\oii\ diagram (Dopita et
al. 2000). \nii/\oii\ is sensitive to metallicity because N is a
secondary element. At high metallicity, \oii\ is suppressed due to the
decrease in temperature. \oiii/\oii\ is a very good proxy of the
ionization parameter. We over-plot the most up-to-date grids for
\hii\ regions from Dopita et al. (2013) with $\kappa=Inf$, which means
the electrons obey the Maxwell-Boltzmann distribution. Our data fall
within the grids. However, when we turn to the BPT diagrams in Figure
\ref{o3hbn2ha.fig} and \ref{o3hbs2ha.fig}, we see that while the high
\ha\ surface brightness spaxels are still consistent with the grid
prediction, DIG dominated low \ha\ surface brightness regions are
located outside the model grids. Specifically, the \nii/\ha\ and
\sii/\ha\ are significantly enhanced in DIG. Previous work found
a similar phenomenon for DIG (e.g., Galarza et al. 1999; Hoopes \&
Walterbos 2003; Kaplan et al. 2016). This means that \hii\ region models
with only a lower $q$ can not explain the low ionization line ratios
like \sii/\ha\ and \nii/\ha. Apart from metallicity and ionization
parameter, there must be other parameter(s) that govern the behavior
of the emission line.

The location of DIG on the BPT diagrams gives us some clue to what the
other parameters could be. DIG enters the composite or LI(N)ER regions
on the \oiii/\hb\ vs \nii/\ha\ and \oiii/\hb\ vs
\sii/\ha\ diagrams. AGN and LI(N)ER are characterized by their hard
ionizing spectrum. A hard spectrum is capable of producing a large
partially-ionized region which enhances the \sii/\ha\ and \nii/\ha,
just like what we observe. Thus, a harder ionizing spectrum is one
possible answer to the line ratios enhancement.

\subsection{Photoionization Grid}
\label{grids.sec}
\begin{figure*}
\includegraphics[scale=0.5]{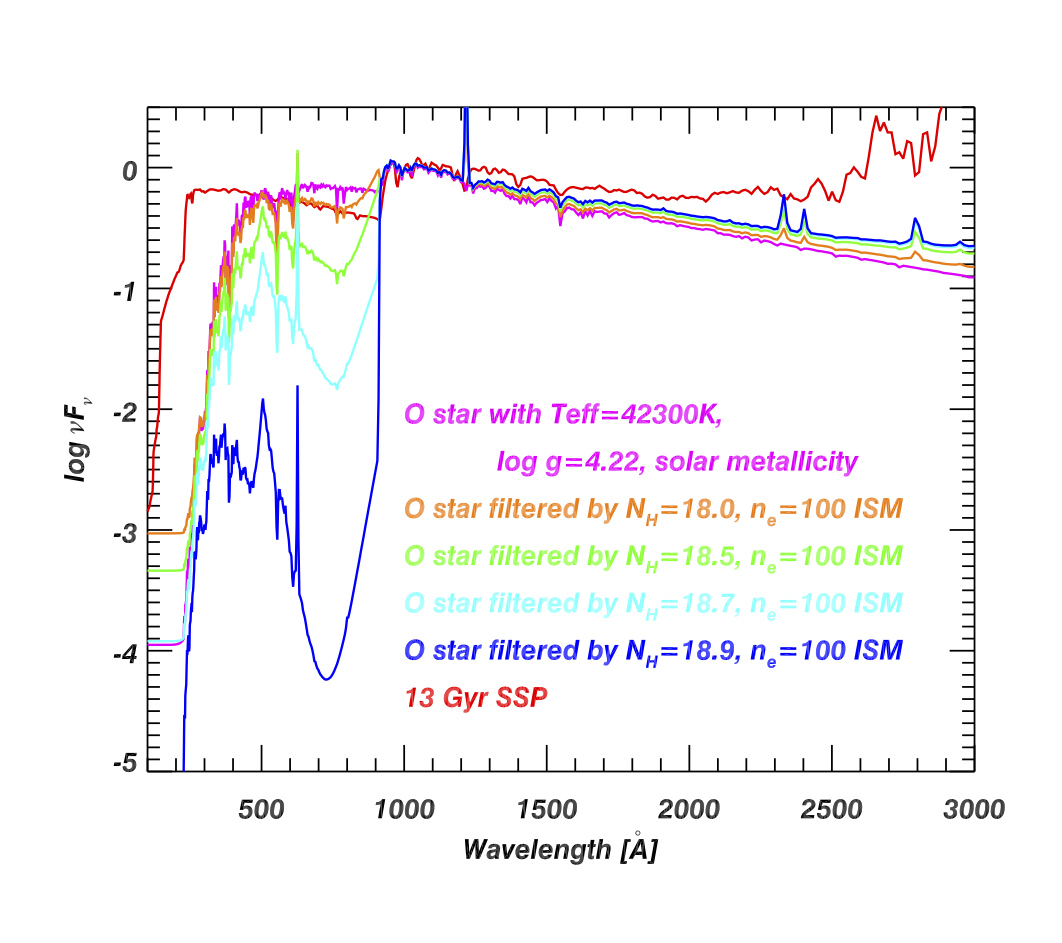} 
\caption{The incident spectra for our models described in Section
  \ref{grids.sec}. All spectra are normalized to the flux at
  1000\AA. The magenta line is a 42,300K, log g=4.22, solar
  metallicity O star generated by Tlusty (Lanz \& Hubeny 2003). The
  orange, green, cyan, and blue lines are the spectra by filtering the
  O star spectrum through column density of $10^{18}$ $cm^{-1}$,
  $10^{18.5}$ $cm^{-1}$, $10^{18.7}$ $cm^{-1}$, $10^{18.9}$ $cm^{-1}$,
  simulating the leaky \hii\ regions. The red line is the spectrum of
  a 13Gyr Simple Stellar Population generated by BC03. }
\label{incident.fig}
\end{figure*}

To get a quantitative idea of how hardness can change line ratios, we
use CLOUDY (Ferland et al. 1998) to calculate new grids for a series
of incident spectra. We test two models that are capable of producing
a hard ionizing spectrum: a leaky \hii\ region model and a hot evolved
star model. The leaky \hii\ region model can have a harder ionizing
spectrum because the lower energy part is more likely to be absorbed
by the IGM (e.g. Giammanco et al. 2005). Hot evolved stars, like pAGB stars, are
characterized by very high temperatures, yielding a hard ionizing
spectrum. The input spectra to test the two models include (see Figure
\ref{incident.fig}):
\begin{description}
\item [(1)] The spectrum of an O star with $T_{eff}=42,300 K$, log g=4.22, solar metallicity. (Tlusty OSTAR2002, Lanz \& Hubeny 2003; magenta line)
\item [(2)] Spectrum in (1) filtered through a gas cloud with $n_e=100$ $cm^{-3}$ and $N_H=10^{18}$ $cm^{-2}$. The resulting spectrum includes both the transmitted spectrum and the emission from the cloud itself. (orange line)
\item [(3)] Spectrum in (1) filtered through a gas cloud with $n_e=100$ $cm^{-3}$ and $N_H=10^{18.5}$ $cm^{-2}$ (green line)
\item [(4)] Spectrum in (1) filtered through a gas cloud with $n_e=100$ $cm^{-3}$ and $N_H=10^{18.7}$ $cm^{-2}$ (cyan line)
\item [(5)] Spectrum in (1) filtered through a gas cloud with $n_e=100$ $cm^{-3}$ and $N_H=10^{18.9}$ $cm^{-2}$ (blue line)
\item [(6)] The spectrum of a 13 $Gyr$ SSP generated with BC03.  (red line)
\end{description}

The metallicity of the ionized cloud is that of Orion Nebula ( Baldwin
et al. 1991; Rubin et al. 1991; Osterbrock et al. 1992; Rubin et
al. 1993). The density of the cloud for model (1) is $10$ $cm^{-3} $,
and $1$ $cm^{-3} $ for model (2), (3), (4), (5) and (6). We run the
calculation for $log$ $U$ =-4.5 to -2.0 with an interval of 0.5~dex
($log$ $U$ =$log$ $q/c$, c is the speed of light).

\begin{figure*}
\includegraphics[scale=0.8]{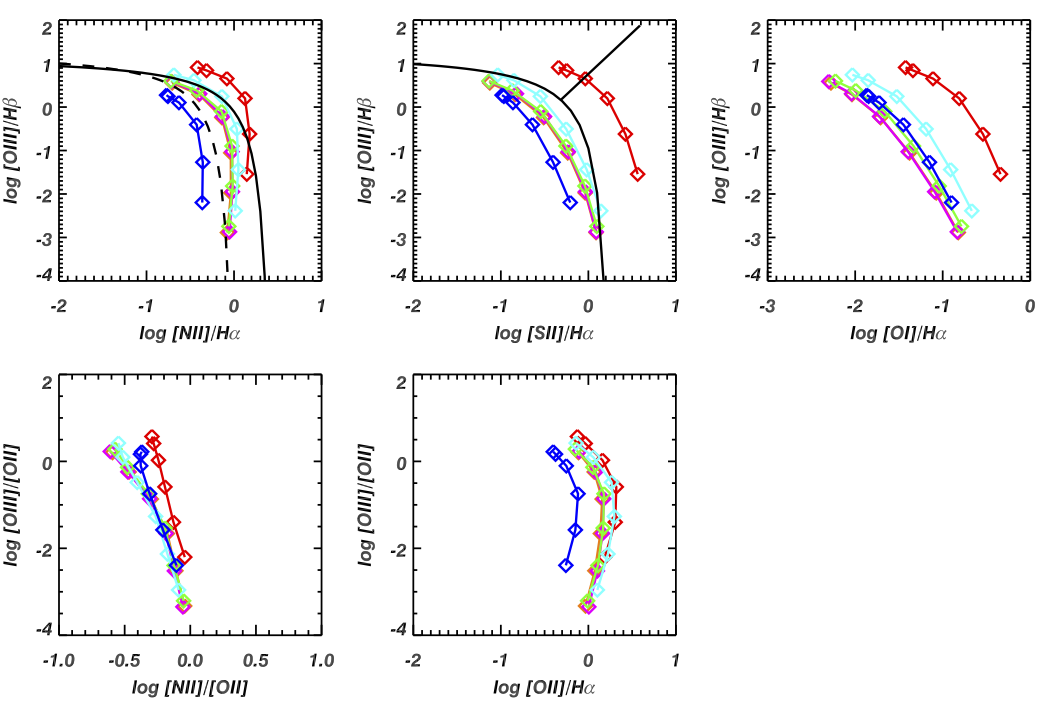} 
\caption{The BPT diagrams, log \oiii/\hb\ vs log \oi/\ha\ diagram, log
  \oiii/\oii\ vs log \nii/\oii\ diagram, and log \oiii/\oii\ vs log
  \oii/\ha\ diagrams for different incident spectra shown in
  Figure~\ref{incident.fig}. The color scheme is the same as in
  Figure~\ref{incident.fig}. The dots with the same color have
  different ionization parameters from log U=-4.5 to log U=-2 with an
  interval of 0.5~dex. The leaky \hii\ region models (orange, green,
  cyan lines) show very similar line ratios to the O star model
  (magenta line) at the same ionization parameter. We note that DIG has
  lower ionization parameters than HII regions. Leaky \hii\ region models can
  not produce enhancement of \sii/\ha\, \nii/\ha\, \oii/\ha\, and
  \oiii/H$\beta$ with a decrease in ionization parameter in most
  cases. The cyan line shows enhancement of these three line ratios,
  but it needs fine tuning of the filtering column density. The blue
  line even shows decreases in \sii/\ha\, \nii/\ha\, \oii/\ha\, and
  \oiii/H$\beta$, because of large \ha\ florescent production, which
  is unphysical. Only ionization by evolved stellar populations (red
  line) can produce enhancement of \sii/\ha\, \nii/\ha\, \oii/\ha\,
  and \oiii/H$\beta$ even when the ionization parameter drops. }
\label{grids.fig}
\end{figure*}

Model (1) represents a typical \hii\ region. Models (2), (3), (4), and
(5) are for testing the leaky \hii\ region model. The hardening of the
ionizing spectrum is obvious when compared with (1).  Model (6)
simulates LI(N)ER-like ionization by an old stellar population which
has been proposed to explain the emission seen in a large fraction of
passive galaxies (e.g., Sarzi et al 2006; Stasi{\'n}ska et al. 2008;
Yan \& Blanton 2012; Kehrig et al. 2012; Singh et al. 2013; Gomes et al. 2016; Belfiore et
al. 2016a,b). Evolved stellar populations have a very hard spectrum.
In Figure~\ref{grids.fig}, we show the diagnostic diagrams for the
different incident spectra listed above. The O star model lies near
the Kauffmann demarcation on the \oiii/\hb\ vs \nii/\ha\ and
\oiii/\hb\ vs \sii/\ha\ diagrams as expected. The 13 $Gyr$ SSP grid
shows significant higher \sii/\ha, \nii/\ha, and \oii/\ha\ relative to
\hii\ regions with the same ionization parameters due to a more
extended partially-ionized region.  Models (2) and (3) show a harder
ionizing spectrum than \hii\ region, but their line ratios are similar
to those for \hii\ regions. Model (4) exhibits enhancement in \nii/\ha,
\sii/\ha, and \oii/\ha\ relative to \hii\ regions. The enhancements are
small, typically less than 0.2~dex. However, model (5), which has the
hardest spectrum blueward of 912\AA\, shows the lowest \nii/\ha,
\sii/\ha, and \oii/\ha. This is because even though those photons
lower than 13.6 $eV$ cannot ionize Hydrogen, they have enough energy
to excite an electron in the Hydrogen atom from ground level (n=1) to
n=3 or n=4 level. When they jump back, they can jump to n=2 which will
produce \ha\ and \hb\ lines. This is fluorescent production. In
model (5), photons pass through a very high column density which absorbs
more than 99.9\%\ of all the ionizing photons. When the transmitted
spectrum is forced to have $log$ $U>-4.5$, we obtain an unphysically
high continuum redward of 912\AA, which leads to large fluorescent
production. It is the enhancement of \ha\ due to fluorescent
production that gives rise to lower \nii/\ha, \sii/\ha, \oii/\ha, and \oi/\ha.

 Considering the large amount of ionizing photons from OB stars
  and the spatial correlation of DIG and \hii\ regions, leaky
  \hii\ regions are likely a major mechanism for producing
  DIG. However, our results disfavor leaky \hii\ region models to
  account for ionization of ALL DIG. DIG has a lower ionization
  parameter than \hii\ regions, a decrease in ionization parameter
  leads to enhancement of \nii/\ha, \sii/\ha, \oii/\ha, and \oi/\ha,
  but a decrease in \oiii/\hb\ at the same time. Leaky \hii\ region
  models can not produce LI(N)ER-like emission. The cyan line shows
enhancement of these four line ratios, but it needs fine tuning of the
filtering column density. DIG that shows LI(N)ER-like emission needs
another ionization source. We favor evolved stars as a major
ionization source for DIG because only ionization by evolved stars
(red line, Model 6) can produce enhancement of \sii/\ha, \nii/\ha,
\oii/\ha, \oi/\ha, and \oiii/\hb\ even when the ionization parameter
drops. Evolved stars are prevalent all over galaxies, and their
contribution to ionizing photons may prevail. Flores-Fajard et
al. (2011) proposed hot low-mass evolved stars (HOLMES) as an
important ionization source for the extra-planar diffuse ionized gas
in edge-on galaxies. For NGC~891, HOLMES begin to contribute more than
50\% of the ionizing photons at scale height |z|>2~kpc. We propose
here that the evolved stars are not only capable of ionizing the
extraplanar gas, but also an important ionizing source for the DIG
near the disk.


\subsection{Photoionization Grids for SSP at different ages}
\label{grids_ssp.sec}
\begin{figure*}
\includegraphics[scale=0.5]{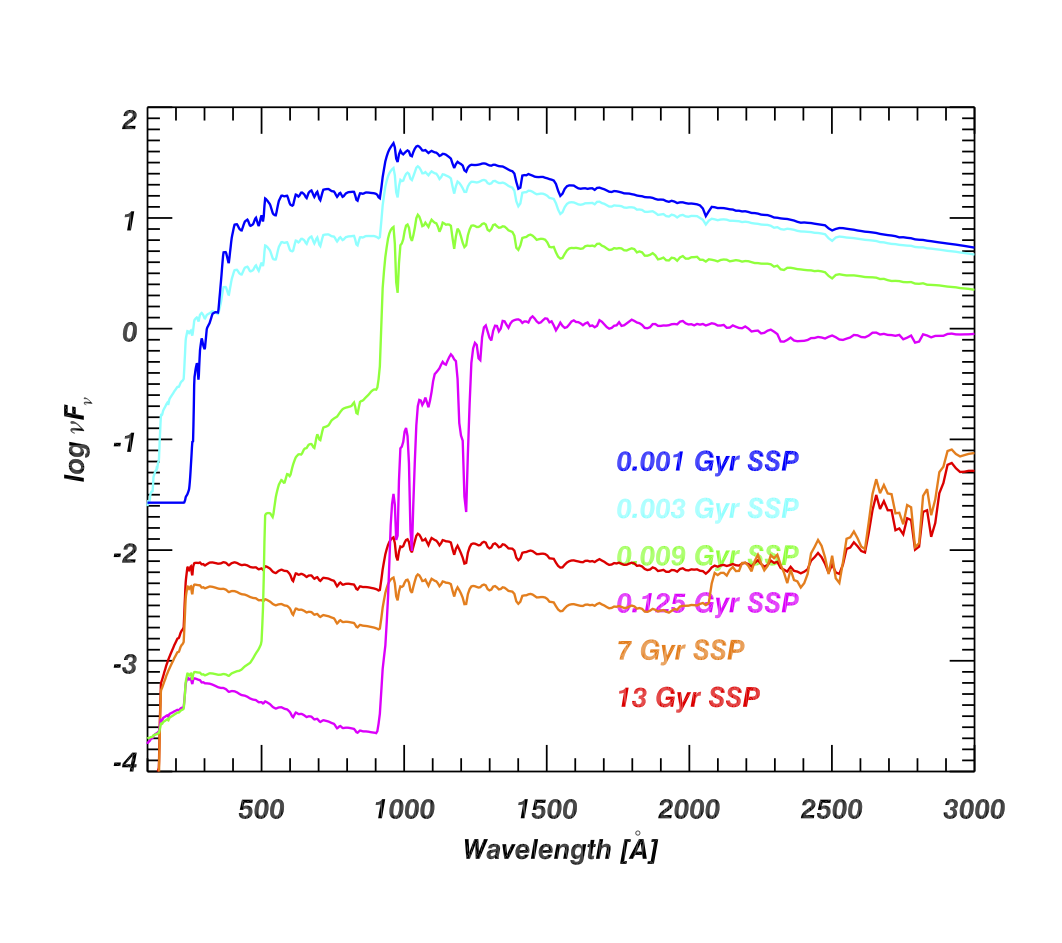} 
\caption{The incident spectra for SSP at 0.001 Gyr, 0.003 Gyr, 0.009 Gyr,
  0.125 Gyr, 7 Gyr, and 13 Gyr with solar metallicity, normalized to flux
  at 6000\AA. We can see that OB stars dominate the ionizing spectrum
  in the beginning, and wane after tens of Myr. After that the
  ionizing spectrum is dominated by emission from evolved stellar
  populations and it hardens. }
\label{incident_ssp.fig}
\end{figure*}

\begin{figure*}
\includegraphics[scale=0.8]{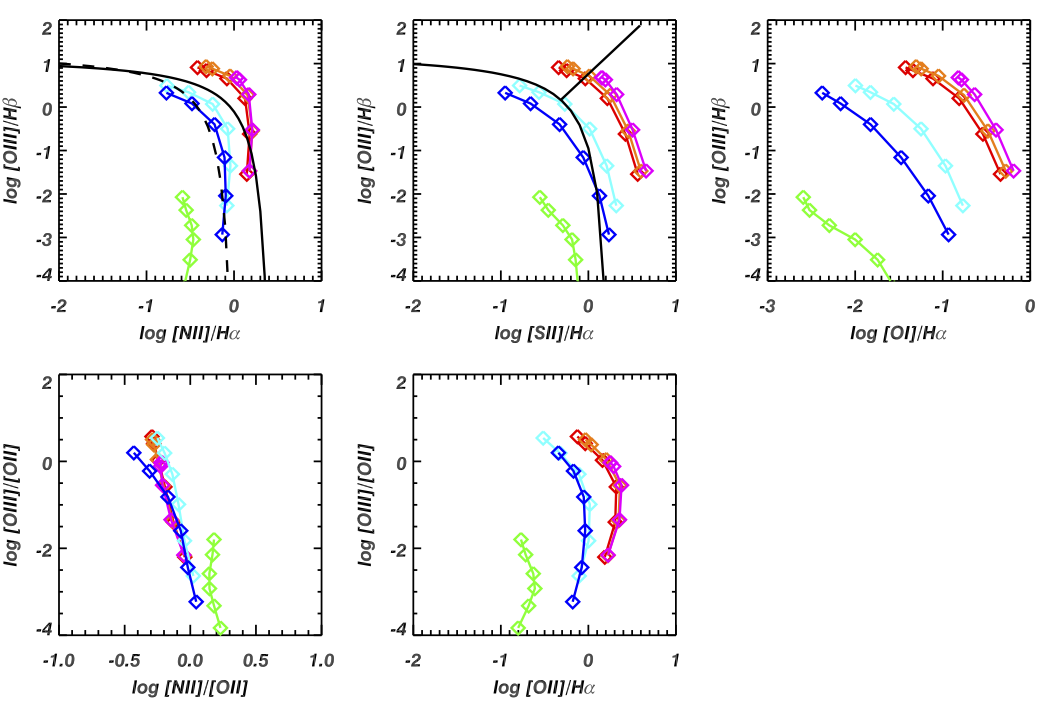} 
\caption{BPT diagrams, log \oiii/\hb\ vs log \oi/\ha, log
  \oiii/\oii\ vs log \nii/\oii, and log \oiii/\oii\ vs log \oii/\ha,
  for different incident spectra of SSPs at different ages. The color
  scheme is the same as Figure~\ref{incident_ssp.fig}. The dots with
  the same color have different ionization parameters from log U=-4.5
  to log U=-2 with an interval of 0.5~dex. At Age<tens of Myr, the
  ionizing spectrum softens as OB stars evolve and die. \sii/\ha\,
  \nii/\ha\, \oii/\ha\, and \oiii/\hb\ all decrease. After the OB
  stars die, the ionizing spectrum is dominated by hot evolved stars,
  and the spectrum hardens. We see increases of \sii/\ha\, \nii/\ha\,
  \oii/\ha\, and \oi/H$\alpha$. This figure shows how the ionized gas
  moves on the diagnostic diagrams as the stellar population ages.  }
\label{grids_ssp.fig}
\end{figure*}

We have shown in last subsection that a 13~$Gyr$ SSP can produce the
LI(N)ER-like emission we see for DIG. One interesting question is: How
would the line ratios change as a SSP ages? To answer this question,
we generate SSPs at 1~$Myr$, 3~$Myr$, 9~$Myr$, 0.125~$Gyr$, 7~$Gyr$,
and 13~$Gyr$ using BC03, and use these spectra as the incident
ionizing spectra for CLOUDY. The spectra from 0.125~$Gyr$ to 13~$Gyr$
change only slightly. The metallicity is that of the Orion Nebula, and
the density is 1~$cm^{-3} $ . We run the calculation for $log$~$U$
=-4.5 to -2.0 with an interval of 0.5~dex. The incident spectra are
shown in Figure~\ref{incident_ssp.fig}. The output line ratios are
plotted in the same color as their incident spectra in
Figure~\ref{grids_ssp.fig}. At 1 $Myr$, the line ratios are located at
the \hii\ region part on the diagnostic diagrams. At 3~$Myr$, the line
ratios are similar to those of 1~$Myr$. At 9~$Myr$, however, we see a
significant decrease in \nii/\ha, \sii/\ha, \oii/\ha, \oiii/\hb,
\oiii/\oii, and an increase in \nii/\oii. This is due to the
significant decrease in 200-500\AA\ spectrum hardness as OB stars age
(Levesque et al. 2010). At 125~$Myr$, the line ratios are already in
the LI(N)ER/AGN region. This means once OB stars die, the line ratios
of the ionized gas would turn into ``LI(N)ER/AGN like'' very fast. The
line ratios do not change very much afterwards. This sheds light on
the interpretation of LI(N)ER/AGN like emission line ratios in
galaxies. When OB stars are alive, their light dominates the ionizing
spectrum and the ionized gas located in the star-forming regions
dominates on the diagnostic diagrams. When OB stars cease to form,
their contribution disappears very fast, and the light of hot evolved
stars begins to dominate after tens of $Myr$, and the ionized gas is
located in the LI(N)ER/AGN regions on the diagnostic diagrams. We have
shown in Section~\ref{grids.sec} that a filtered spectrum can not
reproduce the DIG line ratios with an increase of \oiii/\hb. A higher
\oiii/\hb\ in DIG than in \hii\ regions means the ionization of DIG
may not be linked to OB stars directly but linked to the hot evolved
stars.

We are not claiming that evolved stellar populations are the major
ionizing source of DIG. Overall, the radiative and mechanical energy
from hot evolved stars falls short of the energy budget of diffuse
ionized gas in star-forming galaxies (Reynolds 1984; Ferguson et
al. 1996; Binette et al. 1994). Even if the energy emitted by hot
evolved stars meets the energy requirement of DIG, O stars can provide
at least one order of magnitude more ionizing photons than hot evolved
stars in star-forming galaxies (Reynolds 1984; Ferguson et
al. 1996). Indeed, DIG is found to be linked to \hii\ regions both
along the disk and in vertical direction (e.g. Ferguson et al. 1996;
Zurita et al. 2000; 2002; Rossa \& Dettmar 2003a,b). In star-forming
galaxies, hot evolved stars cannot compete with O stars in total
ionizing photons production. However, we see in two cases that hot
evolved stars might be a major contributor of ionizing photons.  1) In
low surface brightness regions that are located far away from
\hii\ regions. The density of ionizing photons from the \hii\ regions
has dropped significantly so the hot evolved stars begin to
dominate. One example is in regions at large vertical height
(extra-planar gas at high |z|).  Other low surface brightness regions
that show higher \oiii/\hb\ than \hii\ regions probably belong to this
category as well.  2) In galaxies where OB stars have died:
post-starburst galaxies and quiescent/passive galaxies. Quiescent
galaxies are known to show LI(N)ER emissions (Yan et al. 2006; Sarzi
et al. 2006; Stasi{\'n}ska et al. 2008; 
Kehrig et al. 2012; Singh et al. 2013; Gomes et al. 2016; Belfiore et
al. 2015, 2016a,b). We predict DIG to be prevalent in post-starburst
galaxies, and they will show LI(N)ER/AGN like emission.

\section{Impact of DIG on Metallicity Measurement}
\label{bias.sec}
There are several methods to measure gas-phase metallicity in
star-forming galaxies using strong lines in the optical, such as
N2=\nii/\ha, \rtt =(\oiil+\oiiil+\oiiir )/\hb,
O3N2=((\oiii/\hb)/(\nii/\ha)) (e.g., Alloin et al. 1979; Zaritsky et
al. 1994; Pilyugin, 2001; Denicol\'{o} et al. 2002; Pettini \& Pagel,
2004; Pilyugin \& Thuan, 2005) and N2O2=(\nii/\oii) (Dopita et
al. 2000; Kewley et al. 2001), N2S2\ha=8.77+log \nii/\sii\ + 0.264
$\times$ log\nii/\ha\ (Dopita et al. 2016). If the line ratios of DIG
are different from these in \hii\ regions, the metallicity measurement
will inevitably be biased because these metallicity indicators are a
combination of several line ratios. The bias depends on the fraction
of emission contributed by DIG, which increases towards low surface
brightness regions.

\subsection{N2=\nii/\ha}
\label{z_n2.sec}
If we use N2 to derive the metallicity, the enhancement seen in DIG
means that the metallicity will be overestimated in those spaxels with
a high DIG fraction.  This is especially important for observation of
galaxies at high redshift when only the emission lines near \ha\ are
available. We see a typical \nii/\ha\ enhancement of 0.2~dex. We use
the equation in Pettini \& Pagel (2004) to convert \nii/\ha\ to
metallicity: $12+log(O/H)=9.37+2.03\times N2+1.26\times
N2^2+0.32\times N2^3$, N2=log(\nii/\ha). Assuming a \hii\ region has
N2=-0.5, Z(N2)=8.63 in a local galaxy, we obtain Z(N2)=8.865 for DIG
with the same metallicity because N2=-0.3. The typical metallicity
gradient for a local star-forming galaxy is about -0.1~dex
${R_e}^{-1}$ (S{\'a}nchez et al. 2014; Ho et al. 2015), meaning
metallicity drops by $\sim$0.2 from the center to 2$R_e$. The bias due
to DIG would flatten the metallicity gradient from -0.1~dex
${R_e}^{-1}$ to 0 if DIG totally dominates at 2$R_e$.  If we assume a
metal poor \hii\ region has N2=-1.5, Z(N2)=8.08 in a high redshift
galaxy, DIG with the same metallicity has N2=-1.3 and we would obtain
Z(N2)=8.157. The metallicity gradient will be flattened as well. N2 is
usually used for measuring metallicity and metallicity gradients at
high redshift (e.g., Wuyts et al. 2016). Our result indicates that the
presence of DIG potentially flattens N2 derived metallicity gradients
at high redshift.

\subsection{N2O2=\nii/\oii\ }
\label{n2o2.sec}

\begin{figure*}
\includegraphics[scale=0.55]{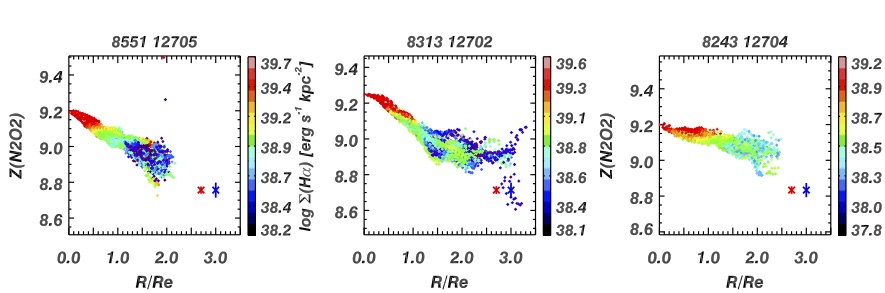} 
\caption{Metallicity derived using N2O2 as a function of
  radius. Z(N2O2) drops with radius due to a metallicity gradient. We
  see that the derived metallicity does not depend on the \hasb\ at a
  fixed radius, and the dispersion at a fixed radius is similar to the
  measuring error. Red and blue bars are errors at log \hasb=39
  \sbunit\ and log \hasb=38.5 \sbunit\ respectively. The length of the
  error bars is 2$\sigma$. We show in Section~\ref{rainbow_z.sec} that
  the impact of DIG is prevalent in all star-forming galaxies in our
  sample, not confined to the three galaxies we show.  }
\label{z_n2o2.fig}
\end{figure*}

N2O2 is a good metallicity indicator because it is not sensitive to
ionization parameter or ionizing spectrum hardness but to metallicity
(Dopita et al. 2000, 2013; Kewley et al. 2002). The caveat of N2O2 is
that it relies on the Nitrogen-to-Oxygen abundance ratio
(N/O). Nitrogen and $\alpha$ elements have different enrichment
timescales, hence their ratio depends on the star formation history
and several other parameters (e.g. Vincenzo et al. 2016) as well as
mixing issues (e.g. Belfiore et al. 2015), especially for low mass
galaxies and galaxy outskirts, where there are prominent variations of
N/O vs O/H relative to metal rich systems and central regions.
Moreover, these diagnostics are sensitive to metallicity only at
12+log(O/H)$>$8.3, i.e. where N/O is a strong function of metallicity,
but they become essentially insensitive to metallicity at
12+log(O/H)$<$8.3, i.e. in the regime where N/O is nearly constant. We
plot metallicities derived using N2O2=\nii/\oii\ as a function of
effective radius and color-code the dots with \ha\ surface brightness
in Figure~\ref{z_n2o2.fig}. N2O2 is converted to metallicity using
$log(O/H) + 12 = $log $[1.54020 + 1.26602\times R + 0.167977\times
  R^2] + 8.93$, R=log \nii/\oii\ (Dopita et al. 2013).  At fixed
radius, Z(N2O2) in the low surface brightness bin is similar to or a
little bit higher than that in the high surface brightness bin.  As
stated in Section \ref{PIR.sec}, \nii/\oii\ only depends on the N/O
abundance ratio and temperature (Dopita et al. 2000, 2013). It is not
subject to the ionization parameter and ionizing spectrum shape
variation.  This makes it an excellent metallicity indicator. The 
 weak dependence of \nii/\oii\ on \ha\ surface brightness at fixed
radius (Figure~\ref{z_n2o2.fig}) demonstrates it is a good metallicity
indicator even in the presence of DIG. However, we note that N2O2
  is subject to variation of N/O ratio (P{\'e}rez-Montero \& Contini 2009; P{\'e}rez-Montero et al. 2013, 2016), temperature variation, and
  uncertainty of extinction correction. Z(N2O2) could be enhanced or
  suppressed for DIG in some regions, maybe due to N/O variation and
  temperature variation. The measurement of extinction involves an
accurate measurement of \hb\ which is not an easy task since the
stellar continuum shows deep \hb\ absorption. This is even harder when
the emission line is weak and S/N is low, such as in the outskirts of
galaxies. Besides, \ha\ and \hb\ arise from the fully ionized region,
while \nii\ and \oii\ are from the partially-ionized region. The
extinction derived using \ha\ and \hb\ may not necessarily be the same
as the extinction experienced by the partially-ionized region. From
Figure~\ref{hba.fig}, we don't see any signs of deviation of
\hb/\ha\ for different \ha\ surface brightness spaxels, so the result
that DIG have similar N2O2 as \hii\ region is robust. We see in
  Figure~\ref{hba.fig} that \hb/\ha\ shows a gradient with positive
  slope even though it does not depend on \hasb. The gradient in
  \hb/\ha\ will change the Z(N2O2) gradient slope. To correct this, we
  suggest getting an \hb/\ha\ gradient by fitting the \hb/\ha\ of
  individual spaxels and apply the overall extinction gradient to
  \nii/\oii\ to get the correct metallicity gradient. In this way, we
  avoid the very noisy \hb/\ha\ when \hasb\ is low when correcting
  extinction. 

A key for probing abundance is to use lines from different elements
with the same ionization potential.  The ionization potential for
$H^0$, $O^{0}$, $O^{+}$, $O^{++}$, $S^{+}$ and $N^{+}$ are 13.598eV,
13.618eV, 35.121eV, 54.936eV, 23.337eV, and 29.601eV respectively.
Ionization potential of \oii\ and \nii\ are similar: 35.121eV and
29.601eV. This makes \nii/\oii\ insensitive to ionization parameter,
thus a good metallicity indicator. The complexity in using
\nii/\oii\ is the extinction correction and the N/O ratio variation at
fixed O/H.


To quantify the change of the metallicity measurements as a function
of \ha\ surface brightness, we do a linear regression fit to the
derived metallicities vs \ha\ surface brightness relation in a narrow
annulus. The distribution of slopes for different metallicity
indicators at three radii are shown in black, blue, and red in
Figure~\ref{rainbow_z.fig}. The metallicity is independent of surface
brightness. The reason we are seeing a dependence of metallicity on
\ha\ surface brightness is due to contamination by DIG.  For Z(N2O2),
the slope distribution is narrow, and peaks at $\sim$0.05. The bias in
metallicity using N2O2 is small and the error is small because N2O2
does not vary much among DIG. DIG has marginally lower Z(N2O2),
suggesting marginally higher temperature.

\subsection{\rtt}
\label{r23.sec}
\begin{figure*}
\includegraphics[scale=0.55]{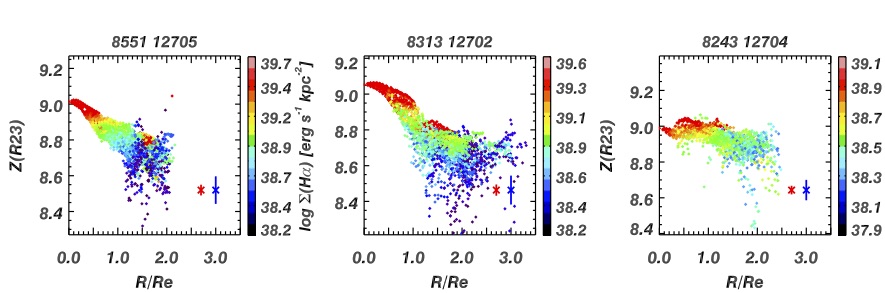} 
\caption{Metallicity derived using \rtt\ as a function of
  radius. Z(\rtt) drops with radius due to metallicity gradient. At
  fixed radius, Z(\rtt) decreases with \hasb, indicating Z(\rtt) is
  systematically biased low for DIG. The measuring errors at log
  \hasb=39 \sbunit\ and log \hasb=38.5 \sbunit\ are indicated using
  red and blue bars. The dispersion of Z(\rtt) is about twice that of
  Z(N2O2). The metallicity gradient would be systematically biased by
  -0.1~dex ${R_e}^{-1}$ if the bias in DIG is not accounted for. }
\label{z_r23.fig}
\end{figure*}

\rtt =(\oiil+\oiiil+\oiiir )/\hb\ is a commonly used metallicity
indicator (McGaugh 1991; Zaritsky et al. 1994; Pilyugin 2001; Kewley
\& Dopita 2002; Kobulnicky \& Kewley 2004).  The advantage is it
does not depend on the N/O ratio. The disadvantage is that the metallicity
is double$-$valued at a given \rtt. In order to use \rtt\ to measure
Z, one has to first determine the ionization parameter and which
branch it is on. Therefore, it always needs to be used in conjunction with
\oiii/\oii\ and N2O2 (or similar).  We use the method of Kewley \&
Ellison (2008) to determine which branch a spaxel is on and derive the
metallicity.  Figure~\ref{z_r23.fig} show how the Z derived from
\rtt\ changes with radius and \ha\ surface brightness.  One can see
that low SB regions have lower measured Z(\rtt) values.  We have
shown in Section~\ref{PIR.sec} that DIG has higher \oii/\hb, lower
\oiii/\oii, and on average similar \oiii/\hb\ as \hii\ regions.  The
combination of the three line ratios leads to the result that Z(\rtt)
is biased to be lower in DIG.

For the whole sample, the Z(\rtt) vs \hasb\ slope distribution peaks at
$\sim$0.2 and the dispersion is large. The contribution of DIG would not
only bias the metallicity measurement systematically, but also
introduce large metallicity measurement uncertainties, as the scatter in the
metallicity vs. radius relation is large.  We note that the gas-phase
metallicity of star-forming galaxies shows a universal gradient of
-0.1~dex ${R_e}^{-1}$. Meanwhile, the slope of Z(\rtt) vs \hasb\ is
$\sim$0.2 . According to Figure~\ref{z_r23.fig}, \hasb\ of DIG
at 2$R_e$ is about 38~\sbunit. Compared to a
\hasb=39~\sbunit\ \hii\ region, the Z(\rtt) bias for DIG would be
0.2. Note the drop of metallicity from the center to 2$R_e$ is only
0.2~dex. The metallicity gradient would be -0.2~dex ${R_e}^{-1}$ if
the outskirt of the galaxy is pure DIG. In other words, the
metallicity measured using Z(\rtt) is biased systematically by
$\sim$-0.1~dex ${R_e}^{-1}$.

\subsection{O3N2}
\label{o3n2.sec}
\begin{figure*}
\includegraphics[scale=0.55]{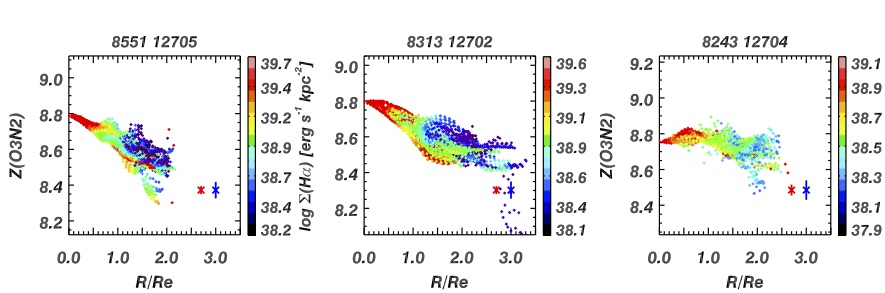} 
\caption{Metallicity derived using O3N2 as a function of
  radius. Z(O3N2) drops with radius due to a metallicity gradient. At
  fixed radius, Z(O3N2) could either decrease or increase with
  \hasb. The measuring errors at log \hasb=39 \sbunit\ and log
  \hasb=38.5 \sbunit\ are indicated using red and blue bars. The
  dispersion of Z(O3N2) is similar to that of Z(N2O2). The metallicity
  gradient would be biased by $\pm$0.05 ${R_e}^{-1}$ if the bias in
  DIG is not considered. }
\label{o3n2.fig}
\end{figure*}

O3N2 is sensitive to oxygen abundance, and it is not impacted by
extinction. The disadvantage of O3N2 is it depends on N/O (P{\'e}rez-Montero \& Contini 2009; P{\'e}rez-Montero et al. 2013, 2016) and the
ionization parameter.  In Figure \ref{o3n2.fig}, we show how DIG would
impact metallicity measurements made from O3N2=(\oiii/\hb)/(\nii/\ha).
We see in Figure \ref{o3n2.fig} that O3N2 is higher in DIG for some
galaxies and lower or similar in other galaxies. The contamination by
DIG could be responsible for a substantial portion of the scatter in
metallicity measurements. When confined only to the high surface
brightness regions, the metallicity gradient derived using O3N2 is
similar to the ones using \rtt\ or N2O2. To make robust metallicity
gradient measurements, one has to properly isolate \hii\ regions and
correct for DIG contamination.

Because the surface brightness generally decreases towards large
radii, the metallicity gradient derived using O3N2 at large radii
might be flatter or steeper than that derived using N2O2 if we include
all the spaxels. For example, in MaNGA galaxy 8313-12702, we get a
metallicity of -0.15~dex ${R_e}^{-1}$ using Z(N2O2). Using only the
high surface brightness regions (red dots) and Z(O3N2) gives the same
result. However, if we include everything, the metallicity gradient
derived using O3N2 will be -0.1~dex ${R_e}^{-1}$. The bias is
significant at least for this galaxy. Z(O3N2) is biased in the other
direction for MaNGA galaxy 8603-12704. Similarly, Mast et al. (2014)
found that the presence of DIG would bias the abundance gradient when
using O3N2 to derive metallicity. The strength of the bias depends a
lot on the assumed calibrator.  We note the degree of bias is not the
same for all galaxies, and we definitely can give a robust metallicity
gradient measurement using high surface brightness regions. The
$\sigma$ of the distribution of Z(O3N2) vs \hasb\ slope at fixed
radius is $\sim$0.1. This translates to a metallicity gradient
dispersion of $\pm$ 0.05~dex ${R_e}^{-1}$ through a similar analysis
as in Section~\ref{r23.sec}, consistent with the different biases we
see in different galaxies. For a large sample of galaxies (S{\'a}nchez
et al. 2014, Ho et al. 2015), the bias for different galaxies may
cancel out when deriving average metallicity gradients.

\subsection{N2S2\ha}
\label{z_N2S2ha.sec}
\begin{figure*}
\includegraphics[scale=0.55]{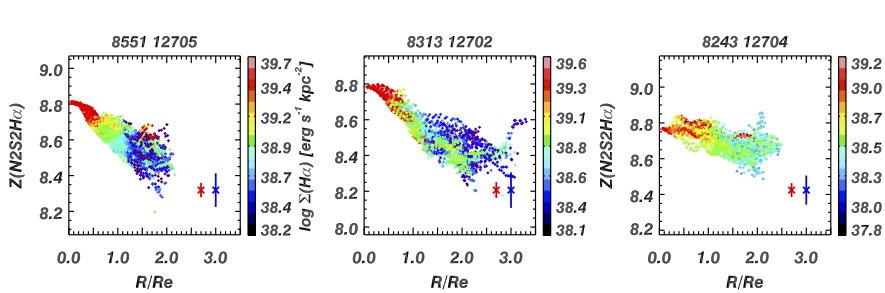} 
\caption{Metallicity derived using \nii/\sii\ and \oii/\ha\ proposed by
  Dopita et al. (2016) as a function of radius. We see that the
  derived metallicity does not depend on the \hasb\ at a fixed radius,
  and the dispersion at a fixed radius is similar to the measuring
  error (red and blue bars are errors at log \hasb=39 \sbunit\ and log
  \hasb=38.5 \sbunit\ respectively). Z(N2S2\ha) performs very similar
  to Z(O3N2). }
\label{z_n2s2ha.fig}
\end{figure*}

Dopita et al. (2016) proposed a new metallicity proxy using only
\niil, \siil, and \ha: 12 + log (O/H)=Z(N2S2\ha)=8.77+log \nii/\sii\ +
0.264 $\times$ log\nii/\ha . It is especially suitable for metallicity
measurements of high redshift galaxies whose spectral wavelength
coverage is limited because the 4 lines used are close to each
other. This estimator is almost linear up to an abundance of 12 + log
(O/H) = 9.05. The caveat is the calibration of this metallicity proxy
is based on a well defined N/O and O/H relation. This calibration
fails for any systems deviating from the assumed N/O$-$O/H
relation. We explore how DIG biases the metallicities derived using
N2S2\ha\ in Figure~\ref{z_n2s2ha.fig}. At fixed radius, we don't see a
significant offset in Z(N2S2\ha) between DIG and \hii\ regions. This
is because Z(N2S2\ha) is insensitive to variation of ionization
parameter and spectral hardness (Dopita et al. 2016), very similar to
N2O2.  Compared with Z(N2O2), the dispersion of Z(N2S2\ha) is
larger. This is understandable because: \\ (a) Four lines are used in
Z(N2S2\ha) while only 2 lines are used to derive Z(N2O2). The
measurement errors enter into the derivation of metallicities. \\ (b) The
relative abundances of N, S, and O are involved in Z(N2S2\ha) while
Z(N2O2) only include N/O. \\ According to the upper right panel of
Figure~\ref{rainbow_z.fig}, The distribution of Z(N2S2\ha) vs
\hasb\ slope is very similar to that of Z(O3N2) in its centroid and
dispersion. The slope distribution for Z(N2S2\ha) has a wider wing,
meaning there are some sources that show significant biases in
Z(N2S2\ha) for DIG and \hii\ regions. This is similar to the scatter
introduced by DIG on O3N2.

\subsection{IZI}
\begin{figure*}
\includegraphics[scale=0.55]{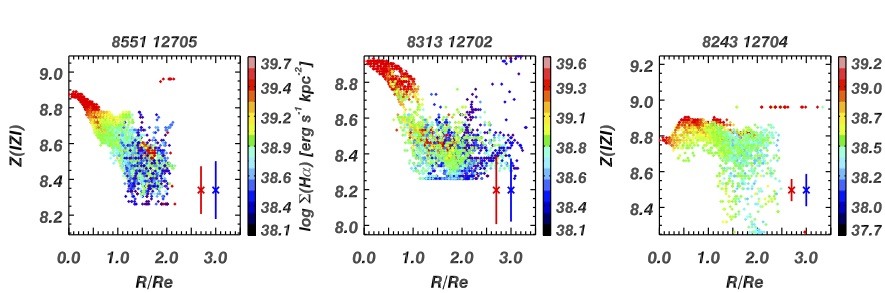} 
\caption{Metallicity derived using IZI by Blanc et al. (2015) as a
  function of radius. The dots are color-coded by \hasb. For high
  \hasb\ regions, the metallicity gradients are similar to those
  derived using other metallicity indicators. However, at fixed
  radius, different surface brightness regions have large metallicity
  discrepancies. For low \hasb\ regions, the metallicities derived from
  IZI can vary by 0.2~dex themselves. This demonstrates that IZI is
  not robust to derive the metallicity of DIG. Note the metallicity
  lower limit is set to 8, so the cut-off at Z$\sim$8.3 is not due to
  a high metallicity lower limit. }
\label{z_izi.fig}
\end{figure*}

IZI (Blanc et al. 2015) uses strong nebular emission lines to derive
the Bayesian posterior probability density function for metallicity
and ionization parameter based on a series of \hii\ region models. In
Figure~\ref{z_izi.fig}, we plot the metallicities derived using IZI as
a function of radius.  We input \oiil, \oiiil, \oiiir, \ha,
  \niir, and \siil\ for the calculation. All the emission line fluxes
  are corrected for reddening using the Balmer decrement
  \ha/\hb\ assuming a Milky Way extinction curve (Fitzpatrick
  1999). We use the values derived with the "output joint mode" in
  IZI. The dots are color-coded by \hasb. For high \hasb\ regions,
IZI gives similar metallicity gradients as the other metallicity
indicators. However, at a fixed radius, different surface brightness
regions (different colors in the plots) have large metallicity
discrepancies. For low \hasb\ regions, the metallicities derived from
IZI can vary by 0.2~dex among themselves. This is at least partly
  because IZI currently only adopts \hii\ region grids. These models
make an inherent assumption that the ionizing spectrum shape is fixed
by the temperature of the OB stars, which is determined by the
metallicity. However, for DIG this assumption does not hold. We have
shown that metallicity+low q for an \hii\ region model can not produce
the line ratios we see in DIG. Thus, IZI is very vulnerable to
contamination by DIG because currently it only contains \hii\ region
models.  However, if one were to include a DIG model in IZI and other Bayesian codes 
to derive metallicities (e.g.,HII-CHI-mistry:  P{\'e}rez-Montero 2014; BOND: Vale Asari et al 2016) , then they 
may be able to give accurate estimates of metallicity and ionization
parameter even with contamination of DIG. 

\begin{figure*}
\includegraphics[scale=0.55]{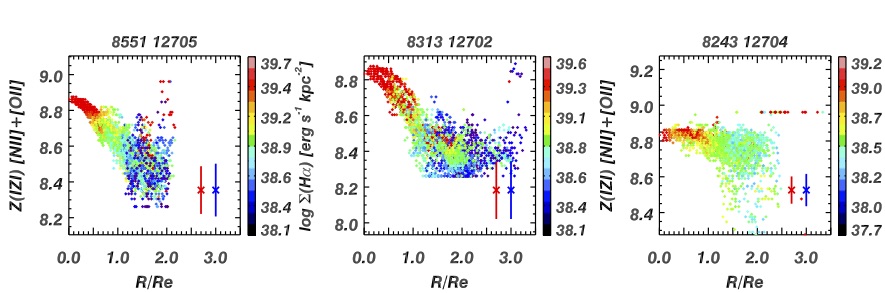} 
\caption{Metallicity derived using IZI by Blanc et al. (2015) as a
  function of radius using only extinction corrected \nii\ and
  \oii. The impact of DIG is much less than when using all strong emission
  lines available. The red and blue bars show the typical line ratio
  errors at log\hasb=39 \sbunit\ and log\hasb=38.5 \sbunit\ for
  individual spaxels. The length of the error bars is 2$\sigma$.  }
\label{z_izi_n2o2.fig}
\end{figure*}

 Furthermore, we input only extinction$-$corrected \nii\ and
  \oii\ into IZI to see if IZI could give a less biased metallicity in
  DIG. \nii\ and \oii\ can only be combined to N2O2 to estimate
  metallicity. Figure~\ref{z_izi_n2o2.fig} shows that the
  metallicity derived through IZI using \nii\ and \oii\ only is better
  than metallicity derived using all strong emission lines
  available. The dispersion at a fixed radius is smaller and the
  dependence on \hasb\ at a fixed radius is weaker. However, Z(N2O2)
  in Figure~\ref{z_n2o2.fig} still shows a tighter metallicity
  gradient than Z(IZI) \nii+\oii. This is because we do not apply
  extinction corrections to individual spaxels when calculating
  Z(N2O2). IZI metallicity with extinction$-$uncorrected \nii\ and
  \oii\ input alone gives very similar result as Z(N2O2). 


\begin{figure*}
\includegraphics[scale=0.55]{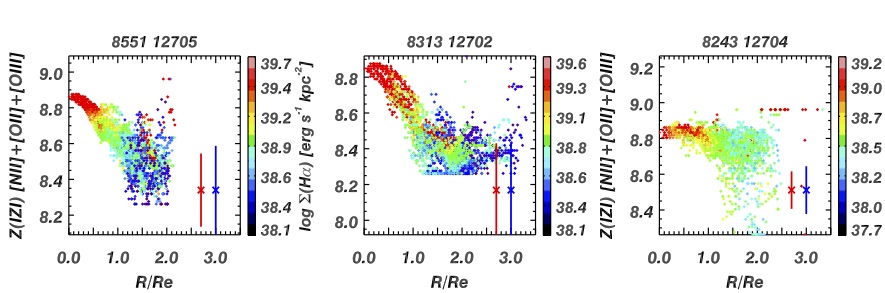} 
\caption{Metallicity derived using IZI by Blanc et al. (2015) as a
  function of radius using only extinction corrected \nii, \oii, and
  \oiii. The result is indistinguishable from the result using only
  \nii\ and \oii. The impact of DIG is much less than when using all
  strong emission lines available. The red and blue bars show the
  typical line ratio error at log\hasb=39 \sbunit\ and log\hasb=38.5
  \sbunit\ for individual spaxels. The length of the error bars is
  2$\sigma$.  }
\label{z_izi_n2o2o3.fig}
\end{figure*}

\subsection{Comparison of different metallicity estimators}
\label{rainbow_z.sec}

\subsubsection{Metallicities vs \hasb\ relation: The Whole Sample}
\begin{figure*}
\includegraphics[scale=0.5]{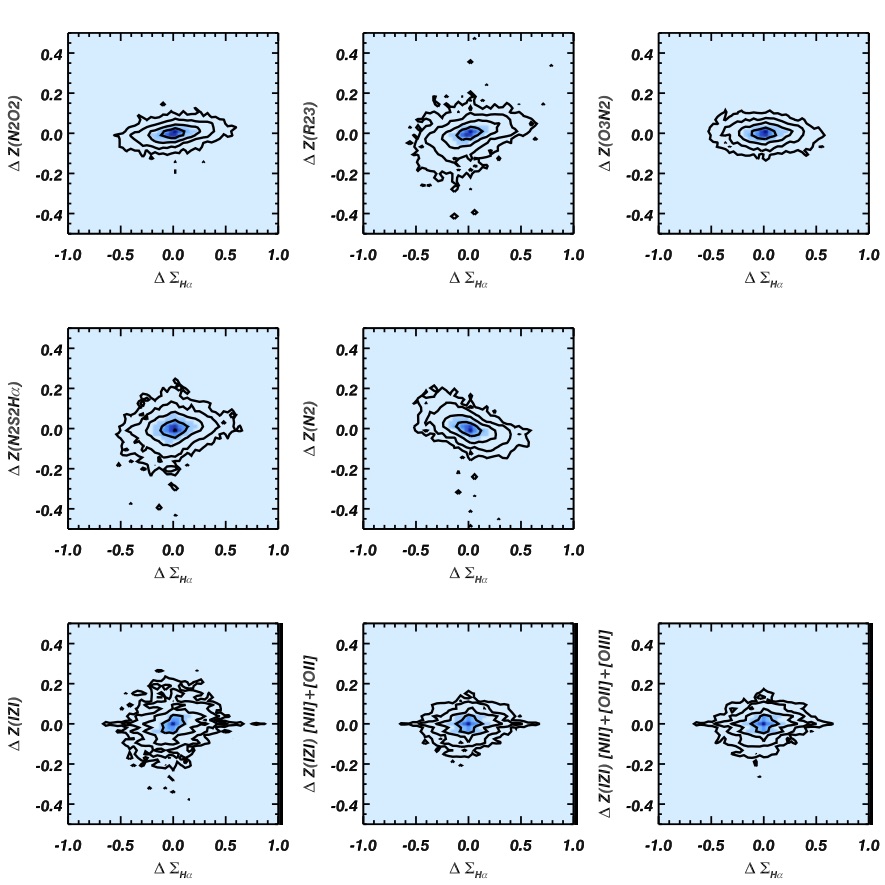} 
\caption{ $\Delta$ Z vs $\Delta$ log \hasb\ relation at [0.4$R_e$,
    0.6$R_e$] for all galaxies. We normalize the $\Delta$ Z vs
  $\Delta$ log \hasb\ relation by subtracting the median Z and median
  log \hasb. All the spaxels in a galaxy are weighted by
  $\frac{1}{N}$, where N is the number of valid spaxels in the
  galaxy. A non-zero slope of the relation means metallicities are
  biased by DIG contribution, and the dispersion of this relation in y direction
  reflect how reliable a metallicity proxy is for an individual galaxy. }
\label{rainbow_z_all.fig}
\end{figure*}

 To quantify the bias introduced by DIG on strong$-$line
  metallicity measurements, we examine the derived metallicities as a
  function of \hasb\ for the whole sample. We select all spaxels in
  [0.4$R_e$, 0.6$R_e$], subtract from all metallicity and log
  \hasb\ measurements their respective medians, then plot all galaxies
  together in the same plot. The spaxels in each galaxy are weighted
  by $\frac{1}{N}$, where N is the number of valid spaxels in this
  galaxy. Figure~\ref{rainbow_z_all.fig} shows the weighted $\Delta$ Z
  vs $\Delta$ log \hasb\ relation for the whole sample. A non$-$zero
  slope of the relation means metallicities are biased by DIG contribution, and the
  dispersion of this relation in y direction reflects how reliable a
  metallicity proxy is for an individual galaxy. The $\Delta$Z(N2O2) vs
  $\Delta$\hasb\ relation has a slope of $\sim$0.05 . The slope of
  $\Delta$Z(\rtt) vs $\Delta$ log \hasb, $\Delta$Z(O3N2) vs $\Delta$
  log \hasb, $\Delta$Z(N2S2\ha) vs $\Delta$log \hasb,, $\Delta$Z(N2)
  vs $\Delta$ log \hasb, relations are $\sim$0.2, -0.05, $\sim$0, and
  -0.2. We define the dispersion as the interval between 90\%
  percentiles contours at $\Delta$ log \hasb =0. For Z(N2O2), the
  dispersion is 0.2, and for Z(\rtt), Z(O3N2), Z(N2S2\ha) and Z(N2),
  the dispersions are 0.35, 0.23, 0.37 and 0.25. Based on slopes and
  dispersions, Z(N2O2) is optimal because the slope is mild and the
  dispersion is smallest. Z(\rtt) has the most significant slope and
  large dispersion. Z(O3N2) has smallest slope and a dispersion only
  slightly larger than Z(N2O2). Z(N2S2\ha) has a nearly 0 slope but
  the dispersion is the largest.  Z(N2) is derived using \nii/\ha,
  thus very vulnerable to DIG contamination. 

\begin{figure*}
\includegraphics[scale=0.5]{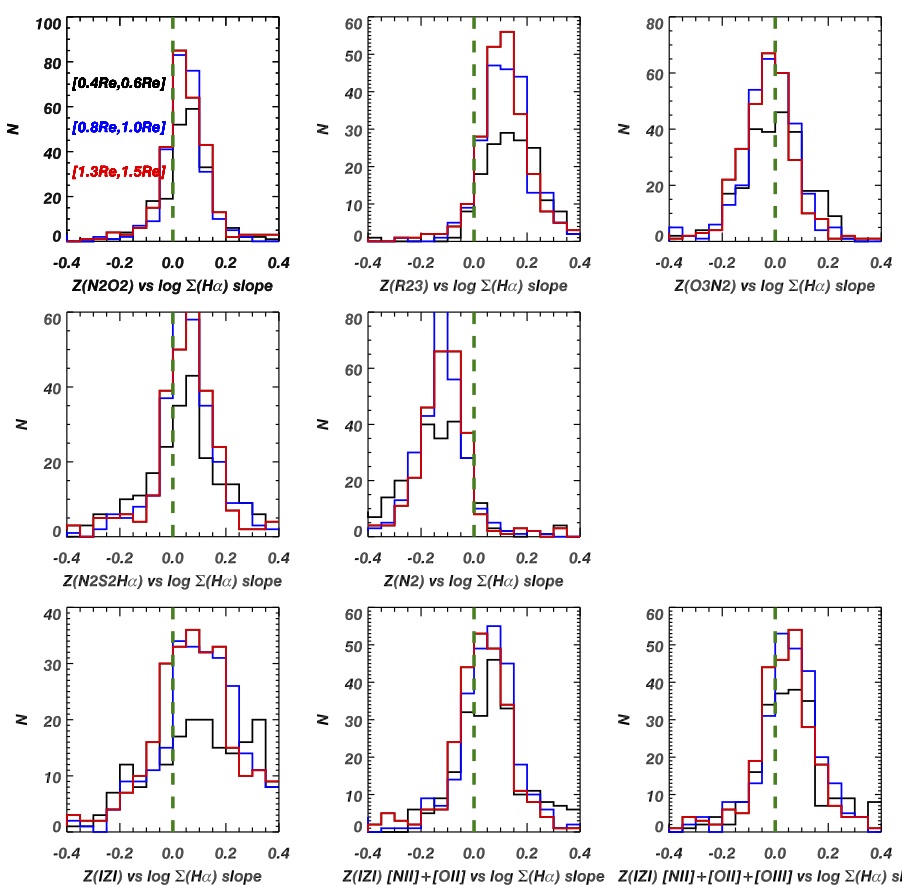} 
\caption{The distribution of metallicities vs $\Sigma(H\alpha)$ slope
  at three radii. The black line is for radius within [0.4$R_e$,
    0.6$R_e$], the blue line is for radius in [0.8$R_e$, 1.0$R_e$],
  and the red line is for the radius in [1.3$R_e$, 1.5$R_e$]. The
  green dashed line has slope=0 line for reference.The metallicity is
  independent of surface brightness; the reason we are seeing a
  dependence of metallicity on \ha\ surface brightness is due to
  contamination by DIG. }
\label{rainbow_z.fig}
\end{figure*}

We also do linear regression to the derived metallicities vs log
\hasb\ relation in narrow radial annuli so the metallicity is
essentially fixed. The distribution of slopes for different
metallicity indicators at three radii are shown in black, blue, and
red in Figure~\ref{rainbow_z.fig}. The intrinsic metallicity is
independent of surface brightness. The reason we are seeing a
dependence of derived metallicities on \ha\ surface brightness is due
to contamination by DIG. A slope of 0 means the DIG does not impact
the metallicity measurements, and a positive slope means DIG biases
the metallicity measurement low. The centroid of the distribution
reflects the systematic offset of that metallicity indicator for the
whole sample, while the dispersion of the distribution reflects the
metallicity error it brings for individual galaxies.  For Z(N2O2), the
slope distribution is narrow, and peaks at
$\sim$0.05~${dex}^{-1}$. The bias in metallicity using N2O2 is small
and the error is small because N2O2 does not vary much among DIG. DIG
regions have marginally lower Z(N2O2), suggesting slightly higher
temperature.  The slope distribution of Z(N2S2\ha) is very similar to
that of Z(N2O2), suggesting Z(N2S2\ha) to be a robust metallicity
estimator even in the presence of DIG.  For Z(\rtt), the slope
distribution peaks at 0.2~${dex}^{-1}$ and dispersion is large. The
contribution of DIG would not only bias the metallicity measurements
systematically, but also introduce a large metallicity measurement
error. This can be seen in the tightness of metallicity vs radius
relation.  For Z(O3N2), the distribution of slope is broad, and peaks
at $\sim$ - 0.05~${dex}^{-1}$. The magnitude of the systematic
metallicity measurement bias is similar for Z(O3N2) and Z(N2O2), but
Z(O3N2) in DIG varies much more significantly.  Using Z(O3N2) will not
bias the measurement if we average a large sample of galaxies, but it
will inevitably bias the metallicity in individual galaxies as shown
in Section~\ref{o3n2.sec}. The large variation of Z(O3N2) and Z(\rtt)
in DIG is expected because these two indicators involve ionization
parameters. For an \hii\ region, this does not matter because the
metallicity and ionization parameter are linked to each other. This
assumption does not hold for DIG. The variation in the ionization
parameter in DIG enters the metallicity measurement. The impact of DIG
on using Z(N2), Z(\rtt) and Z(O3N2) metallicity gradients is
discussed in Section~\ref{z_n2.sec}, Section~\ref{r23.sec} and
Section~\ref{o3n2.sec} respectively.  For Z(IZI), the metallicities
derived for DIG are systematically lower. Besides, the dispersion of
the slope distribution is broadest, with $\sigma \sim$0.3. This means
IZI not only systematically biases the metallicity of DIG lower, but
also introduces large errors into the metallicity measurements.

\subsubsection{Metallicities vs \hasb\ relation: Split by Stellar Mass}
\begin{figure*}
\includegraphics[scale=0.55]{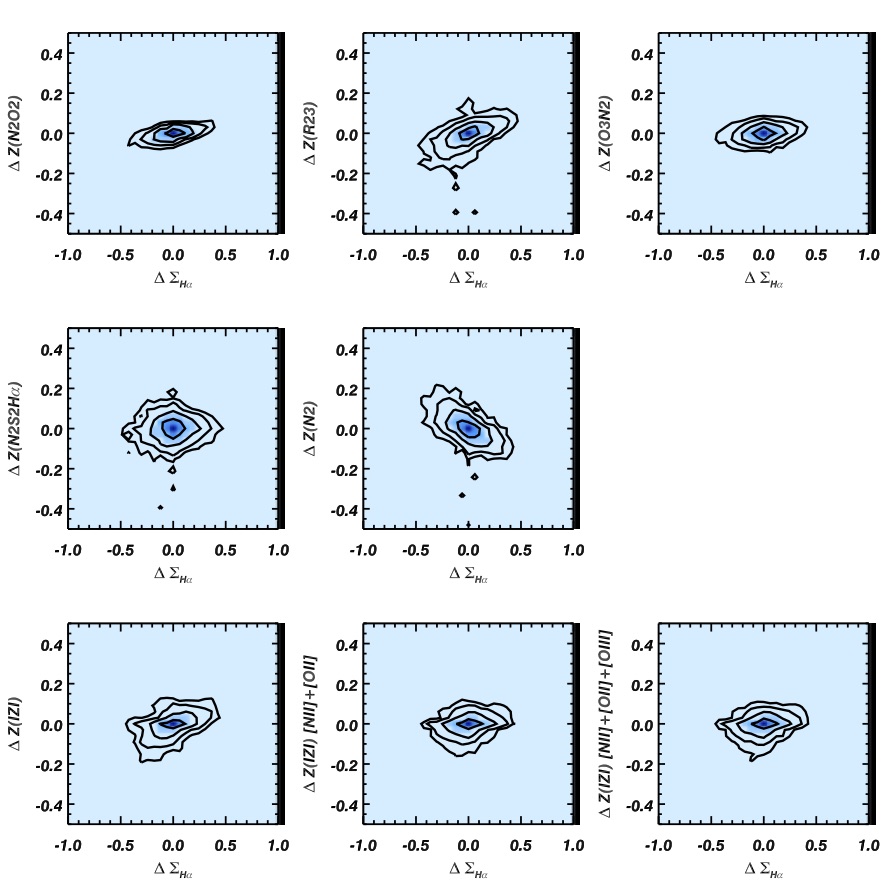} 
\caption{ $\Delta$ Z vs $\Delta$ log \hasb\ relation at [0.4$R_e$,
    0.6$R_e$] for the one third most massive galaxies in our
  sample. We normalize the Z vs log \hasb\ relation by subtracting the
  median Z and median log \hasb. All the spaxels in a galaxy are
  weighted by$\frac{1}{N}$, where N is the number of valid spaxels in
  the galaxy.  }
\label{rainbow_z_Mstarhigh.fig}
\end{figure*}

\begin{figure*}
\includegraphics[scale=0.55]{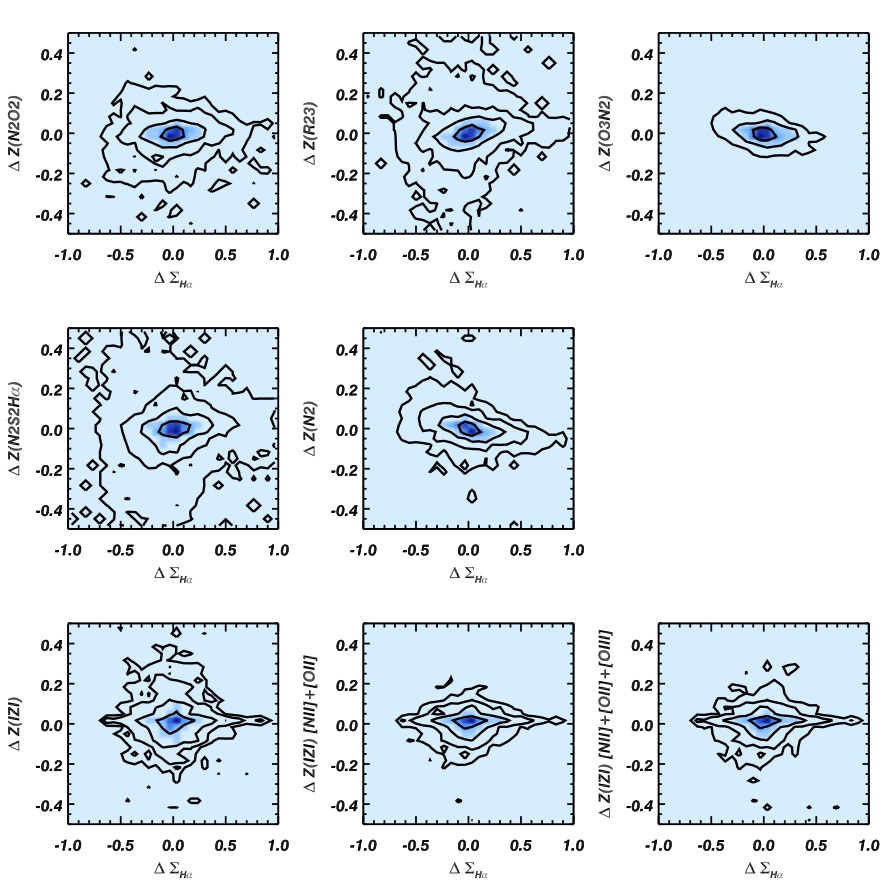} 
\caption{ $\Delta$ Z vs $\Delta$ log \hasb\ relation at [0.4$R_e$,
    0.6$R_e$] for the one third least massive galaxies in our
  sample. We normalize the Z vs log \hasb\ relation by subtracting the
  median Z and median log \hasb. All the spaxels in a galaxy are
  weighted by$\frac{1}{N}$, where N is the number of valid spaxels in
  the galaxy.  }
\label{rainbow_z_Mstarlow.fig}
\end{figure*}

 In Figure~\ref{rainbow_z_Mstarhigh.fig} and
  Figure~\ref{rainbow_z_Mstarlow.fig}, we show the normalized Z vs \hasb\ relations for galaxies with stellar mass less
  than $10^{9.43}$ and higher than $10^{10.08}$ respectively. They are
  the one third least massive and one third most massive galaxies in
  our sample. The $\Delta$Z(N2O2), $\Delta$Z(\rtt), $\Delta$Z(N2S2\ha) and
  $\Delta$Z(N2) vs $\Delta$ log \hasb\ relationa are similar in the
  most massive and least massive galaxies. However, the
  $\Delta$Z(O3N2) vs $\Delta$ log \hasb\ relations in the most massive
  galaxies have positive slopes while those in the least massive
  galaxies have negative slopes. This is mostly due to the dependence
  of the $\Delta$ log \oiii/\hb\ vs $\Delta$ log \hasb\ relation on
  stellar mass. The dependence of DIG's impact on stellar mass means
  it is crucial to take care of DIG contamination when comparing the
  metallicity and metallicity gradient of galaxies of different masses
  using Z(O3N2).

In summary, 
\begin{enumerate}
\item Metallicities derived using N2O2 are optimal because they
  exhibit the smallest bias and error.
\item Metallicities derived using the O3N2 or N2S2\ha\ (Dopita et
  al. 2016) for DIG can be significantly higher or lower than those
  for \hii\ regions. Using O3N2 or N2S2\ha\ to derive metallicity can
  bias the metallicity gradient by $\pm$0.05 ${R_e}^{-1}$ for an
  individual galaxy if the contamination by DIG is not accounted
  for. For a large sample of galaxies (S{\'a}nchez et al. 2014, Ho et
  al. 2015), the bias for different galaxies may cancel out when
  deriving average metallicity gradients.
\item \rtt\ derived metallicities for DIG are lower than those for
  \hii\ regions due to a lower ionization parameter. Using \rtt\ to
  derive metallicity will systematically bias the metallicity gradient
  by $\sim$-0.1 ${R_e}^{-1}$ because of DIG.
\item Using N2=\nii/\ha\ to derive metallicities will systematically
  bias the metallicity gradient by $\sim$0.05-0.1~dex ${R_e}^{-1}$,
  considering that DIG typically shows 0.2~dex higher \nii/\ha.
\item IZI works well for the \hii\ region dominated regions, but
  fails for deriving metallicities of DIG, probably because IZI
    currently only contains \hii\ region models. 
\end{enumerate}

\subsection{Metallicity Gradients in the local universe and at high redshift}
For the local universe, we usually have \oii, \oiii, \hb, \nii, \ha,
and \sii\ to derive metallicities. Recent works find that the local
star-forming galaxies exhibit a universal metallicity gradient of
-0.1~dex ${R_e}^{-1}$ within 2 $R_e$ (e.g. S{\'a}nchez et al. 2014; Ho
et al. 2015). These works employ O3N2 (S{\'a}nchez et al. 2014, Ho et
al. 2015) and N2O2 (Ho et al. 2015) to derive the metallicity.
  S{\'a}nchez et al. (2014) isolated \hii\ regions from DIG and used
  O3N2 to derive the metallicities using \hii\ regions. The
  contamination of DIG is small. Ho et al. (2015) rejected the
  DIG-dominated spaxels by imposing S/N$>$3, and Z(O3N2) and Z(N2O2)
  give consistent metallicities and metallicity gradients. The
  metallicity gradient at R$<2R_e$ is robust against DIG
  contamination. Using MaNGA data and the robust metallicity estimator
  N2O2 to remeasure metallicities would be worthwhile to re-examine the
  metallicity gradient at large radius. 

At high-redshift, we need a new calibration of strong line metallicity
indicators because the physical properties are different in high
redshift galaxies compared to local galaxies. High redshift
star-forming galaxies are systematically offset towards higher
\oiii/\hb\ on the BPT diagram relative to the local star-forming
galaxy locus (Shapley et al. 2005; Erb et al. 2006; Liu et al. 2008;
Brinchmann et al. 2008; Hainline et al. 2009; Wright et al. 2010;
Trump et al. 2011 Kewley et al. 2013a,b; Steidel et al. 2014). The
line ratio shifts of high-redshift star-forming galaxies are similar
to DIG. Based on photoionization models, Steidel et al. (2014)
concluded that the offset on the BPT diagram of the z$\sim$2.3
star-forming galaxies is caused by a combination of harder stellar
ionizing radiation field, higher ionization parameter, and higher N/O
at a given O/H compared to most local galaxies. A new calibration
between N2 and O3N2 was obtained. A higher N/O abundance has been
  claimed to explain the offset in the BPT Diagram (Masters et
  al. 2014, Shapley et al. 2015, Sanders et al. 2016). There is
debate wether the metallicity gradients at high redshift are flatter or
steeper (Yuan et al. 2011; Swinbank et al. 2012; Jones et al. 2013,
2015; Leethochawalit et al. 2016; Wuyts et al. 2016). It takes time
and resources to obtain all \oiii, \hb, \nii, \ha\ for a galaxy at
high redshift. In most cases, either \oiii+\hb\ or \nii+\ha\ is
observed. The few lines available limit the metallicity estimator to
either \rtt\ or N2, two indicators that are impacted most by the
presence of DIG. This makes the metallicity measurements at high-z
vulnerable to contamination of DIG. However, we note current studies
  are unlikely to be seriously affected by DIG because in general they
  are forming stars at an incredibly high rate and they are very well
  represented by pure \hii\ region spectra. Future studies at high
  redshift probing low surface brightness limits (e.g., with JWST) may
  have to take into account the contribution of DIG. One needs to be
  cautious when making comparisons of high-z and low-z line ratios
  because DIG matters more at low z. 


\section{Other Relevant Issues}
\label{discussion.sec}

\subsection{Line ratios vs \ha\ surface brightness relation for an individual \hii\ region}
\label{local.sec}
Rela{\~n}o et al. (2010) studied the line ratios and \ha\ surface
brightness of NGC 595, one of the most luminous \hii\ regions in
M33. The scale of NGC 595 is about 300pc. By comparing the
\ha\ surface brightness distribution shown in their Figure~3 and the
\nii/\ha, \sii/\ha, and \oiii/\hb\ maps in Figure~6, we can see that
the low surface brightness regions, which are located far away from
the ionizing stars, have high \nii/\ha, \sii/\ha, and lower
\oiii/\hb. The line ratios vs. \ha\ surface brightness relations are
derived on a scale of hundreds of pc around an individual
\hii\ region. Madsen et al. (2006) saw the same trend in their
observation of Milky Way. The resolution of the WHAM survey is
1~degree, which corresponds to 90~pc at 5~kpc.  So the physical scale
of WHAM is tens to hundred parsecs. Do these local relations on scales
of hundreds of parsecs extend to kpc scale probed by MaNGA? MaNGA's
spatial resolution of 1~kpc means each PSF may include tens or even
hundreds of \hii\ regions. If the \ha\ surface brightness vs line
ratios relation is confined to 300pc scales, when smeared by a
kpc-scale PSF, the line ratio variation will be smoothed out and
disappear. In other words, if DIG is dominated by regions within 10pc
of each \hii\ region, then we would not see the line ratios vs
  \hasb\ relation after smoothing it with a kpc-sized PSF.  {\it The
  smearing effect can produce a surface brightness gradient but can
  not produce an anti-correlation between surface brightness and line
  ratios.} The fact that we see a surface brightness vs line ratio
relation on kpc-scales means that these relations for individual
\hii\ region can not be local, but extend to kpc scales. The
DIG we see is dominated by regions far away (kpc) from individual
\hii\ regions.  In edge-on galaxies, we do see that DIG can extend to
a few kpc above the galaxy plane (e.g, Rand 1996, 1997; Rossa
\& Dettmar 2000; Kehrig et al. 2012; Jones et al. 2016).

Then what does the line ratio vs \ha\ surface brightness relation tell
us? What we need to keep in mind is that we are looking at the average
properties of the galaxy at our resolution. As discussed in detail in
Appendix~\ref{smear.sec}, this relation is preserved for individual
\hii\ regions no matter what resolution we use to observe the
galaxy. However, under poor resolution, many \hii\ regions are mixed,
and their relations are also mixed. We don't know how many
\hii\ regions there are in one spaxel and what their line ratio vs
\ha\ surface brightness relations looks like. What we see is the
average properties within 1~kpc, and how the line ratios and surface
brightness change coherently on this scale. The curvature of the
relation does tell us that an \hii\ region is not a simple
Str{\"o}mgren sphere but includes an extra component of DIG which has
different line ratios. At 1~kpc scale, we can model the ISM as a
combination of DIG and \hii\ regions with their respective line
ratios. This relation also illustrates the feasibility of using
\ha\ surface brightness to roughly separate DIG and \hii\ regions.



\subsection{Shedding Light on Composite Region and LI(N)ER}
Any factors that change the line ratios can influence the positions on
the diagnostic diagrams (e.g., Zhang et al. 2008; P{\`e}rez-Montero \&
Contini 2009).  In Figure \ref{o3hbn2ha.fig} and \ref{o3hbs2ha.fig},
we explored the distribution of each spaxel on the BPT diagram:
\oiii/\hb\ vs \nii/\ha\ and \oiii/\hb\ vs \sii/\ha\ (Baldwin et
al. 1981; Veilleux \& Osterbrock 1987). In the \oiii/\hb\ vs
\nii/\ha\ diagram, we plot the demarcations from Kewley et al. (2001)
and Kauffmann et al. (2003) to identify the ionizing source
properties. We color-code the dots by \ha\ surface brightness, and the
dots in composite \& LI(N)ER region have low \ha\ surface brightness.
As discussed in Section~\ref{diagnostic.sec}, a hardened O star
spectrum filtered by ISM can not produce LI(N)ER-like emissions, while
hot evolved stars like pAGB stars can produce the LI(N)ER-like line
ratios.
In Belfiore et al. (2016a), they classified galaxies that show
LI(N)ER-like emission into two classes: cLIER and eLIER.  cLIER shows
LI(N)ER-like emission in the center, while eLIER shows extended
LI(N)ER-like emission all over the galaxy.  In some galaxies in our
sample, the line ratios could extend to LI(N)ER regions on the BPT
diagrams (See also Galarza et al. 1999; Kaplan et al. 2016). We
suspect these two phenomenon may have a common origin. However, the galaxies that show eLIER emission and cLIER could have
different stellar population and gas metallicity from the sample we
study in this paper. Therefore a lot more work is needed to prove 
if they are really the same. We defer this to a future paper.
The Shocked POststarburst Galaxy Survey (SPOGS, Alatalo et
al. 2016a,b) explores a sample of galaxies selected from SDSS Data
Release 7 that show LI(N)ER like emission line ratios and
post-starburst spectral features. A similar phenomenon is found by Yan
et al. (2006). After OB stars die, the spectrum of the galaxy is
dominated by spectral features of A stars, characterizing the
post-starburst signature. According to our grids, the decrease in OB
stars and increase LI(N)ER-like emission line ratios are naturally
linked through increasing photoionization by evolved stars. DIG,
cLIER, and SPOGs are very similar to each other, and the study of
their similarity and distinctions can greatly help us understand the
ionized gas in different types of galaxies. From our figures, DIG
dominated regions can mimic composite or even AGN spectra. If the
selection of AGN is purely based on the BPT diagram, star-forming
galaxies with significant DIG could be incorrectly classified as
AGN. DIG and AGNs might be distinguished using the WHAN diagram (Cid Fernandes et al 2010, 2011) 
due to their different EW(\ha) and \nii/\ha\ distributions. This will be explored further in the future.   
Our analysis demonstrates the importance of considering the
presence of DIG while making optical line ratio
classifications. Additionally, the analysis illustrates the complexity
caused by DIG in interpreting the line ratio diagnostic diagrams. The
so-called composite galaxies or LI(N)ERs could be coming from a
variety of objects.

 Shocks could also produce LI(N)ER-like line ratios. Shocked gas
  with velocities less than 500~\kms\ occupy the LI(N)ER part of the
  diagnostic diagrams (Farage et al. 2010; Rich et al. 2011), while
  shocked gas with velocities greater than 500~\kms\ fall in the
  Seyfert part of the diagram (Allen et al. 2008; Kewley et
  al. 2013). Shock model can produce the enhanced \sii/\ha, \nii/\ha, 
  \oii/\hb, \oi/\ha\ and \oiii/\hb\ we see. Additionally, the temperature of shocked gas is elevated
  (Allen et al. 2008). To test if shocks are indeed a major ionization
  source for DIG, high spectral resolution spectra ($\sim$4000) are
  needed to separate different velocity components and constrain the
  kinematics. Our spectral resolution (R$\sim2000$) limits our ability
  to test for shocks. We leave the question of the full
  nature of ionization sources of DIG for further studies. 

\subsection{How to minimize the impact of DIG}
The best way to minimize the impact of DIG is to separate DIG and
\hii\ regions spatially. With high spatial resolution IFS data like
CALIFA, it is possible to resolve individual \hii\ regions to lower
the contamination of DIG. For low spatial resolution data like MaNGA
and SAMI, the individual spaxels are covering kpc scale regions, thus
the emission is a mix of \hii\ regions and DIG. The mixing also happens for MUSE data of high redshift galaxies. 
For these data, selecting high \hasb\ spaxels would mitigate the impact of DIG. As we
have shown in this paper, the high \hasb\ regions have \hii\ region
line ratios. Besides, the metallicity gradient derived using only the
high \hasb\ regions for different strong line method: Z(N2O2),
Z(\rtt), Z(O3N2), Z(N2S2\ha) are consistent with each other. This
means the high \hasb\ regions, even though contaminated by DIG due to
beam smearing, are \hii\ region dominated. A \hasb\ cut is a robust
and easy way to reduce the impact of DIG. The exact value of
\hasb\ cut depends on the spatial resolution. For our MaNGA survey,
\hasb$>$39 \sbunit\ select reliable \hii\ region dominated
spaxels. An equivalent width (EW) cut is not recommended since EW depends on
metallicity (e.g., Tresse et al. 1999). For low metallicity regions, for example in low metallicity
galaxies or the outskirt of a galaxy, EW will be high due to the low
metallicity. A EW cut suitable for the center of a galaxy will select DIG contaminated spaxels at the outskirt of this galaxy. 
So selecting high \hasb\ regions is a reliable and convenient way to
minimize DIG contamination.



\section{Summary and Conclusions}
\label{conclusion.sec}
We selected a sample of 365 blue face-on galaxies from 1391 galaxies
observed by MaNGA, and illustrated the impact of DIG on line ratios,
interpretation of diagnostic diagrams, and metallicity
measurements. We find that \ha\ surface brightness is a good indicator
to separate \hii\ regions from DIG. DIG shows distinct properties as
listed below:
\begin{enumerate}
\item $[$S\,{\footnotesize II}$]$/\ha, \nii/\ha, \oii/\hb, and
  \oi/\ha\ are enhanced in DIG relative to \hii\ regions.
\item DIG has lower \oiii/\oii, indicating lower ionization
  parameter. \oiii/\hb\ of DIG can be higher or lower than
  \hii\ regions.
\item On BPT diagrams, contamination by DIG moves \hii\ regions
  towards composite or LI(N)ER-like regions. A harder ionizing
  spectrum is needed to explain DIG line ratios.
\item Leaky \hii\ region models only shift the line ratios slightly
  relative to \hii\ region models, thus fail to explain
  composite/LI(N)ER line ratios displayed by DIG. 
  Leaky \hii\ region models cannot explain the \oiii/\hb\ but do pretty well for the other line ratios.
\item Our result favors ionization by evolved stars as a major
  ionization source for DIG with LI(N)ER-like emission.
\item Metallicities derived using N2O2=\nii/\oii\ are optimal because
  they exhibit the smallest bias and scatter.
\item Metallicities derived using the O3N2=(\oiii/\hb)/(\nii/\ha) or
  N2S2\ha=8.77+log \nii/\sii\ + 0.264 $\times$ log\nii/\ha\ (Dopita et
  al. 2016) for DIG can be significantly higher or lower than those
  for \hii\ regions. Using O3N2 or N2S2\ha\ to derive metallicities can
  bias the metallicity gradient by $\pm$0.05~dex${R_e}^{-1}$ for an
  individual galaxy if the contamination by DIG is not accounted
  for. \rtt\ derived metallicities for DIG are lower than those for
  \hii\ regions due to a lower ionization parameter. Using \rtt\ to
  derive metallicities will systematically bias the metallicity gradient
  by $\sim$-0.1~dex ${R_e}^{-1}$ because of DIG. Using N2=\nii/\ha\ to
  derive metallicities will systematically bias the metallicity gradient
  by $\sim$0.05-0.1~dex ${R_e}^{-1}$, considering that DIG typically
  shows 0.2~dex higher \nii/\ha.
\item The metallicities in high redshift galaxies are mostly derived
  using \rtt\ or N2, rendering their metallicity and metallicity
  gradient measurements most vulnerable to the impact of DIG. Using
  Z(N2S2\ha) for high redshift galaxies is more robust to prevent the
  contamination by DIG. For most of the recent high$-$z observations,
  the contamination by DIG is probably not severe because we only see
  high \hasb\ regions. When comparing the metallicities of high-z
    galaxies with local galaxies, one needs to use caution since
    DIG might impact metallicity measurements of local galaxies. 
\end{enumerate}

\section*{Acknowledgements}
KB is supported by World Premier International Research Center Initiative (WPI Initiative), MEXT, Japan and by JSPS KAKENHI Grant Number 15K17603. 
MAB acknowledges support from NSF AST-1517006.
RM acknowledge support by the Science and Technology Facilities Council (STFC) and the ERC Advanced Grant 695671 "QUENCH?.
C.A.T. acknowledges support from National Science Foundation of the United States grant no. 1412287
DB was supported by grant RSF 14-50-00043.
AD acknowledges support from The Grainger Foundation. 
Funding for the Sloan Digital Sky Survey IV has been provided by the Alfred P. Sloan Foundation, the U.S. Department of Energy Office of Science, and the Participating Institutions. SDSS- IV acknowledges support and resources from the Center for High-Performance Computing at the University of Utah. The SDSS web site is www.sdss.org.
SDSS-IV is managed by the Astrophysical Research Consortium for the Participating Institutions of the SDSS Collaboration including the Brazilian Participation Group, the Carnegie Institution for Science, Carnegie Mellon University, the Chilean Participation Group, the French Participation Group, Harvard-Smithsonian Center for Astrophysics, Instituto de Astrof\'isica de Canarias, The Johns Hopkins University, Kavli Institute for the Physics and Mathematics of the Universe (IPMU) / University of Tokyo, Lawrence Berkeley National Laboratory, Leibniz Institut f\"ur Astrophysik Potsdam (AIP), Max-Planck-Institut f\"ur Astronomie (MPIA Heidelberg), Max-Planck-Institut f\"ur Astrophysik (MPA Garching), Max-Planck-Institut f\"ur Extraterrestrische Physik (MPE), National Astronomical Observatory of China, New Mexico State University, New York University, University of Notre Dame, Observatório Nacional / MCTI, The Ohio State University, Pennsylvania State University, Shanghai Astronomical Observatory, United Kingdom Participation Group, Universidad Nacional Aut\'onoma de M\'exico, University of Arizona, University of Colorado Boulder, University of Oxford, University of Portsmouth, University of Utah, University of Virginia, University of Washington, University of Wisconsin, Vanderbilt University, and Yale University.

\bibliography{bibliography3}
\bibliographystyle{mnras}

\noindent \hrulefill

\noindent 
$^{1}$Department of Physics and Astronomy, University of Kentucky, 505 Rose Street, Lexington, KY 40506, USA \\
$^{2}$Kavli Institute for the Physics and Mathematics of the Universe (Kavli IPMU, WPI), Todai Institutes for Advanced Study, the University of Tokyo, Kashiwa 277-8583, Japan  \\
$^{3}$Department of Astronomy, University of Wisconsin-Madison, 475 N. Charter Street, Madison, WI, 53706, USA  \\
$^{4}$Department of Astronomy, New Mexico State University, Las Cruces, NM 88003, USA \\
$^{5}$Cavendish Laboratory, University of Cambridge, 19 J. J. Thomson Avenue, Cambridge CB3 0HE, United Kingdom \\
$^{6}$Kavli Institute for Cosmology, University of Cambridge, Madingley Road, Cambridge, United Kingdom \\
$^{7}$Institute of Cosmology {\&} Gravitation, University of Portsmouth, Dennis Sciama Building, Portsmouth, PO1 3FX, UK \\
$^{8}$Department of Astronomy, University of Texas at Austin, Austin, TX 78712, USA \\
$^{9}$Max-Planck-Institut f{\"u}r Astrophysik, Karl-Schwarzschild-Str. 1, D-85748 Garching, Germany \\
$^{10}$Instituto de Astronom{\'i}a, Universidad Nacional Autonoma de Mexico, A.P. 70-264, 04510, Mexico, D.F., Mexico \\
$^{11}$Apache Point Observatory and New Mexico State University, P.O. Box 59, Sunspot, NM, 88349-0059, USA \\
$^{12}$Sternberg Astronomical Institute, Moscow State University, Moscow, Russia \\
$^{13}$Unidad de Astronom{\'i}a, Universidad de Antofagasta, Avenida Angamos 601, Antofagasta 1270300, Chile \\
$^{14}$PITT PACC, Department of Physics and Astronomy, University of Pittsburgh, Pittsburgh, PA 15260, USA \\
$^{15}$Department of Physics and Astronomy, University of Utah, 115 S. 1400 E., Salt Lake City, UT 84112, USA \\
$^{16}$Tsinghua Center of Astrophysics \& Department of Physics, Tsinghua University, Beijing 100084, China \\
$^{17}$Shanghai Astronomical Observatory, Nandan Road 80, Shanghai 200030, China \\
$^{18}$Space Telescope Science Institute, 3700 San Martin Drive, Baltimore, MD 21218, USA \\
$^{19}$Departamento de F{\'i}sica, Facultad de Ciencias, Universidad de La Serena, Cisternas 1200, La Serena, Chile \\
$^{20}$Universidade Federal do Rio Grande do Sul, IF, CP 15051, Porto Alegre 91501-970, RS, Brazil \\
$^{21}$Laborat\'orio Interinstitucional de e-Astronomia - LIneA, Rua Gal. Jos\'e Cristino 77, Rio de Janeiro, RJ - 20921-400, Brazil \\

\appendix
\section{Beam Smearing Effect}
\label{smear.sec}
\begin{figure*}
\includegraphics[scale=1.0]{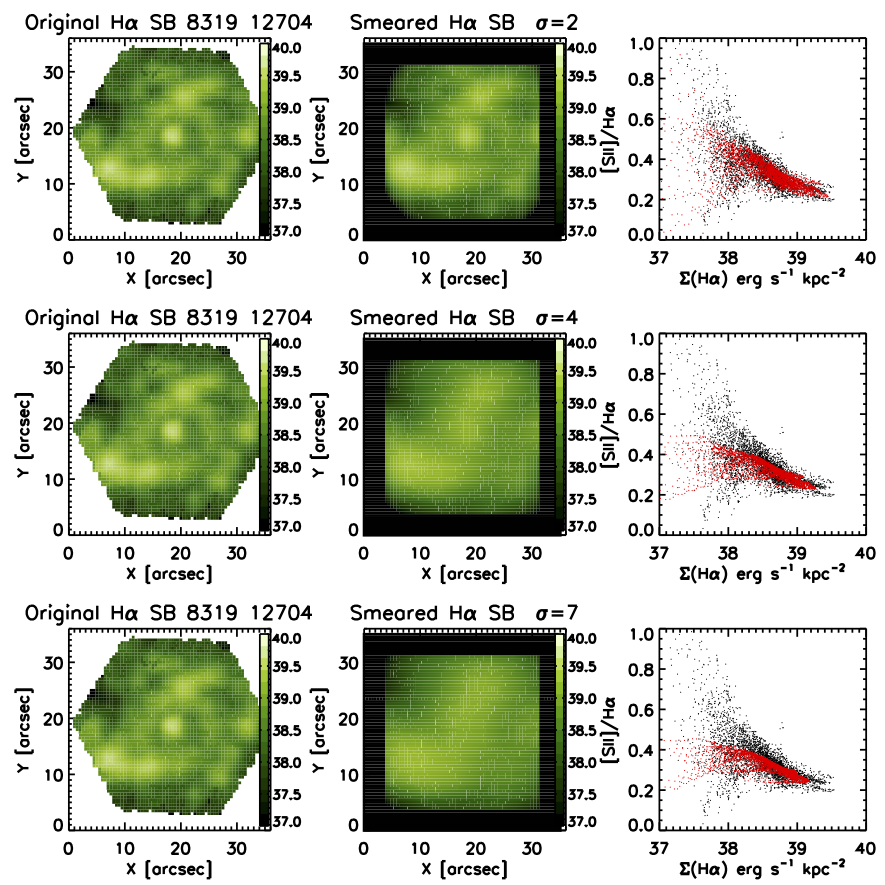} 
\caption{Left panel: the original $\Sigma(\ha)$ map, middle panel: the smeared  $\Sigma(\ha)$ map. Right panel: the comparison of original (black dots) and smeared (red dots) line ratio vs \ha\ surface brightness relation. The \ha\ flux map and \nii\ flux map is convolved with a 2D gaussian with $\sigma$ pixels listed in the middle panels of each row.  }
\label{smear.fig}
\end{figure*}

Most of the  \hii\ reigon is smaller than 1~kpc, and the size depends on the density of the gas (e.g. Hunt \& Hirashita 2009). Under MaNGA resolution, the mixing of \hii\ region with DIG and other \hii\ region is unavoidable. What's the effect of this mixing? We have seen in Section \ref{s2.sec} that in one annulus, the line ratio vs surface brightness relation is tight. What would happen if we mix 2 regions on the line ratio vs surface brightness relation? From an analytical calculation, the mixed point is still on the same relation. However, if we mix two or more \hii\ regions together, what would the line ratio vs \ha\ surface brightness relation like? 
 We show in Figure \ref{smear.fig} for a simulation of observing our galaxies with poorer resolution. The left panel is the original $\Sigma(\ha)$ map, and the middle panel is the smeared  $\Sigma(\ha)$ map. In the right panel, we show the comparison of original (black dots) and smeared (red dots) line ratio vs \ha\ surface brightness relation. The \ha\ flux map and \sii\ flux map is convolved with a 2D gaussian with $\sigma$ pixels listed in the middle panels of each row.   We see several fact from this exercise: 
\begin{enumerate}
\item The beam smearing effect does not change the line ratio vs SB relation for one \hii\ region, so the curve we derive using MaNGA data is likely to preserve when higher resolution data is obtained. In other word, this relation is intrinsic, reflecting the relationship between line ratio and surface brightness. Our resolution of $\sim1$~kpc means one \ha\ flux peak in our map is probably a mixture of several \hii\ regions. \\  
\item It is impossible to disentangle DIG from \hii\ region from the curve alone.  \\
\item The smearing makes the relation tighter and shorter.  \\
\item When mixing two or more \hii\ regions together, their line ratio vs SB relations also merge to an intermediate one.  \\
\end{enumerate}
The smearing process is an averaging process. We are witnessing more and more of the average properties of the \hii\ region and DIG of the galaxy as we go to poorer resolution. The overall smeared line ratio vs \ha\ surface brightness relation is tighter than before. So the tightness of the relation we see should be partly due to the beam smearing effect. 

\bsp	
\label{lastpage}
\end{document}